\documentclass{article}
\usepackage{LaThuileFPSproCenter}
\input{psfig.sty}
\newcommand {\pom} {I\!\! P}
\begin{document}
\title{DIFFRACTION AND EXCLUSIVE (HIGGS?) PRODUCTION FROM CDF TO LHC~\footnote{
~~This paper is composed of updated versions of two papers presented at {\em DIFFRACTION 2006 - International Workshop on Diffraction in High Energy Physics, September 5-10 2006, Adamantas, Milos island, Greece}, published by {\em Proceedings of Science:} PoS (DIFF2006) 016 and PoS (DIFF2006) 044.}\\
\vglue 1em
{\small Presented at {\em Les Rencontres de Physique de la Vall\'{e}e d'Aoste,\,} La Thuile, Aosta Valley, Italy, March 4-10, 2007. }}
\author{Konstantin Goulianos\\
  {\em The Rockefeller University, 1230 York Avenue, New York, NY 10023, USA}\\
}
\maketitle

\baselineskip=11.6pt

\begin{abstract}

The diffractive program of the CDF Collaboration at the Fermilab Tevatron $p\bar p$ Collider is reviewed with emphasis on recent results from Run II at $\sqrt s=$1.96~TeV. Results are presented on the $x$-Bjorken and $Q^2$ dependence of the diffractive structure function obtained from dijet production, on the slope parameter of the $t$-distribution  of diffractive events as a function of $Q^2$ in the range $1\hbox{ GeV}^2<Q^2<10^4\hbox{ GeV}^2$, and on cross sections for exclusive dijet, $e^+e^-$, and $\gamma\gamma$ production. The exclusive dijet and $\gamma\gamma$ production rates are used to check theoretical estimates of exclusive Higgs boson production at the Large Hadron Collider. 
Other data on soft and hard diffraction from $pp/p\bar p$  collisions, and also data from diddractive deep inelastic scattering are presented and interpreted in the RENORM phenomenological model, in which cross sections are obtained from the underlying inclusive parton distribution function of the nucleon and QCD color factors. 
\end{abstract}
\newpage

\section{Experimental results\label{experiment}}
\subsection{Introduction}
The CDF collaboration has been carrying out a systematic and comprehensive program of studies of diffractive interactions since the start of operations of the Fermilab Tevatron $p\bar p$ collider in 1989. The ultimate goal of this program is to provide experimental results which will be of help in elucidating the QCD character of hadronic diffraction~\cite{HCP2006}. Diffractive interactions are characterized by large rapidity gaps~\footnote{Rapidity gaps are regions of rapidity devoid of particles; rapidity, $y=\frac{1}{2}\frac{E+p_L}{E-p_L}$, and pseudorapidity, $\eta=-\ln\tan\frac{\theta}{2}$, are used interchangeably, as in the kinematic region of interest the values of these two variables are approximately equal.} in the final state, presumed to occur via the exchange of a quark/gluon combination carrying the quantum numbers of the vacuum. This exchange is traditionally referred to as {\em Pomeron}~\cite{books}. 
The process which is directly analogous to the classical diffraction of light is elastic scattering, but it is inelastic diffraction processes that provide the most stringent tests for QCD inspired models of diffraction. The total cross section is also of interest in testing theoretical models of diffraction, since it is related to the imaginary part of the forward elastic scattering amplitude through the optical theorem. In this paper, we present results obtained at the Tevatron by CDF and comment on their physics significance.

The names/dates of the Tevatron runs and integrated luminosities of data collected by CDF are listed below:  

\begin{center}
\noindent\begin{tabular}{llr}
Run Number&Date&$\int$Lum (pb$^{-1}$)\\
\hline
\underline{Run~I}&\\
I\O&1988-1989&5\\
Ia&1992-1993&20\\
Ib&1993-1995&100\\
Ic&1995-1996&10\\
\underline{Run II}&\\
IIa&2003-2006&1000\\
IIb&currently in progress&\\
\end{tabular}
\end{center}

In Run~I\O\,, CDF measured elastic, single diffractive, and total cross sections at $\sqrt s=$630 and 1800 GeV. 

\begin{figure}[ht]
\centerline{\psfig{figure=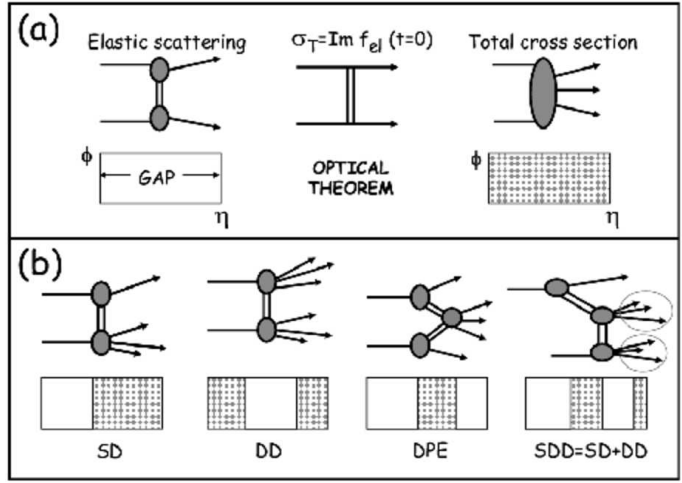,width=0.75\textwidth}}
\caption{Schematic diagrams and event topologies in azimuthal angle $\phi$ vs. pseudorapidity $\eta$ for (a) elastic and total cross sections, and (b) single diffraction (SD), double diffraction (DD), double Pomeron exchange (DPE), and double plus single diffraction cross sections (SDD=SD+DD). The hatched areas represent regions in which there is particle production.}
\label{fig:diagrams}
\end{figure}

In Runs~Ia, Ib and Ic, CDF studied both soft and hard diffractive processes, with the latter incorporating a hard partonic scattering in addition to the characteristic large rapidity gap of diffraction. Fig.~\ref{fig:diagrams} shows schematically the diagrams and final state event topologies of the processes studied by CDF in Run~I.  

In Run~II, the CDF diffractive program was enhanced by extending the kinematic range of the measurements of hard diffractive processes and by additional studies of exclusive production processes. 

All Run~I results have been published. These results are briefly summarized. The Run~II results are discussed in more detail. Recent results on the $x$-Bjorken and $Q^2$ dependence of the diffractive structure function and on the $t$-dependence of diffractive cross sections are reported and characterized in terms of their physics content. In addition, results on exclusive dijet, $e^+e^-$, and $\gamma\gamma$ production are presented and their significance in calibrating predictions for exclusive Higgs boson production at the LHC is discussed. 
%
\subsection{Run I\O\, Results}
In Run I\O\,, CDF measured the elastic, soft single diffractive, and total $p\bar p$ cross sections at $\sqrt s=$630 and 1800 GeV.  
The measurement was performed with the CDF~I detector, which during run I\O\, had tracking coverage out to $|\eta|\sim 7$ and Roman Pot Spectrometers on both sides of the Interaction Point (IP). The normalization was obtained by the luminosity independent method, which is based on simultaneously measuring the total interaction rate, which depends on $\sigma_T$, and the elastic scattering differential rate at $t=0$, which depends on $\sigma_T^2$ (optical theorem):
$$
\sigma_T \propto \frac{1}{L} \left(N_{el}+N_{inel}\right)\;\;\;\;\;\&\;\;\;\;\; \sigma_T^2\sim\frac{1}{1+\rho^2}\frac{dN_{el}}{dt}|_{t=0}$$
$$\Rightarrow\;\;\;\; 
\sigma_T=\frac{16\pi}{1+\rho^2}\frac{1}{N_{el}+N_{inel}} \frac{dN_{el}}{dt} |_{t=0}
$$
Paradoxically, overestimating the total rate, as for example due to background events, yields smaller elastic and total cross sections, while loss of inelastic events results in larger cross sections. 

\begin{figure}[h]
\phantom{xxx}
\vglue 3em
{\hspace*{-4em}\psfig{figure=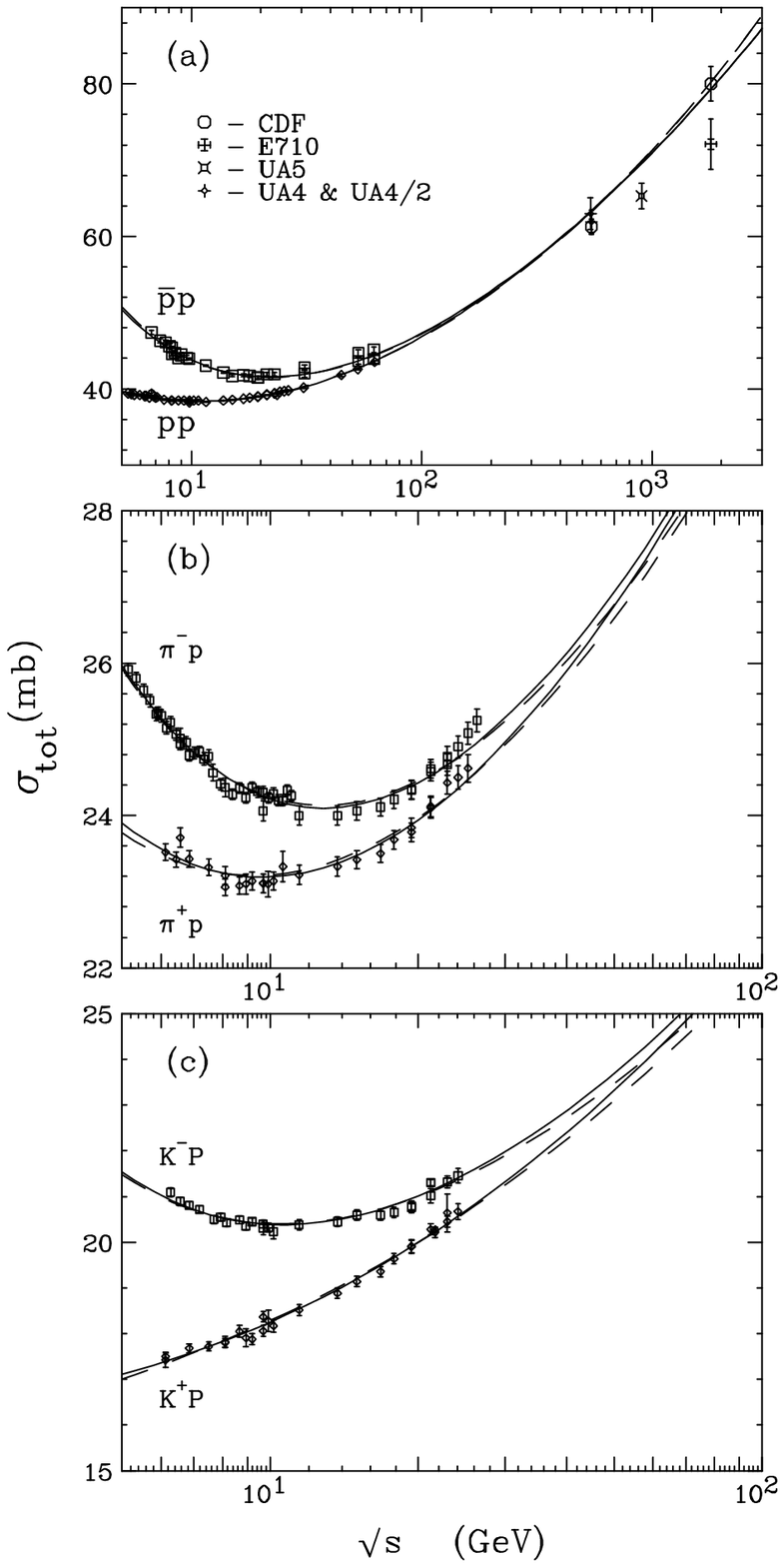,width=0.6\textwidth}\hspace*{-13em}\psfig{figure=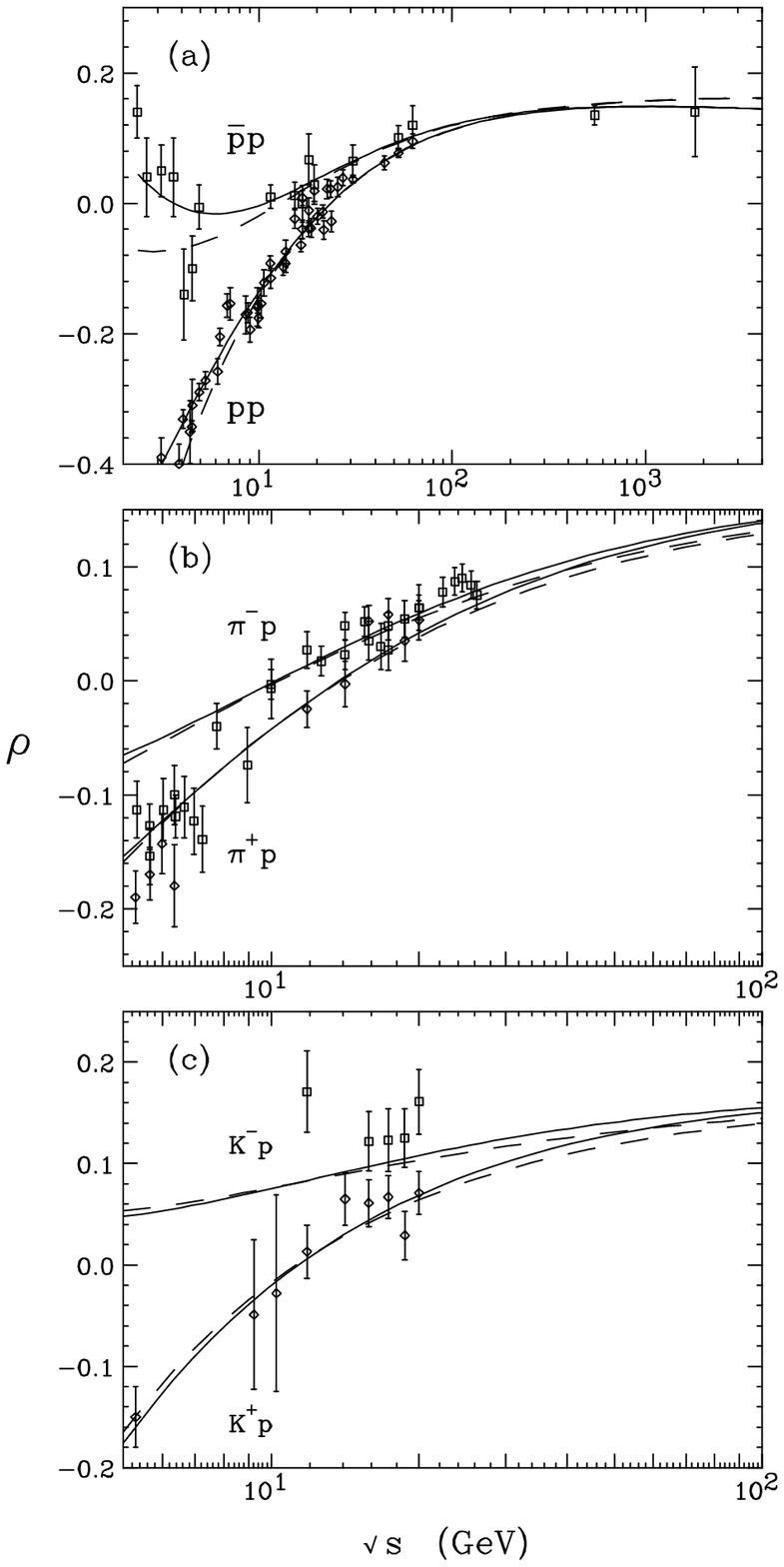,width=0.6\textwidth}}
\vglue -14em
\hspace*{18em}{\psfig{figure=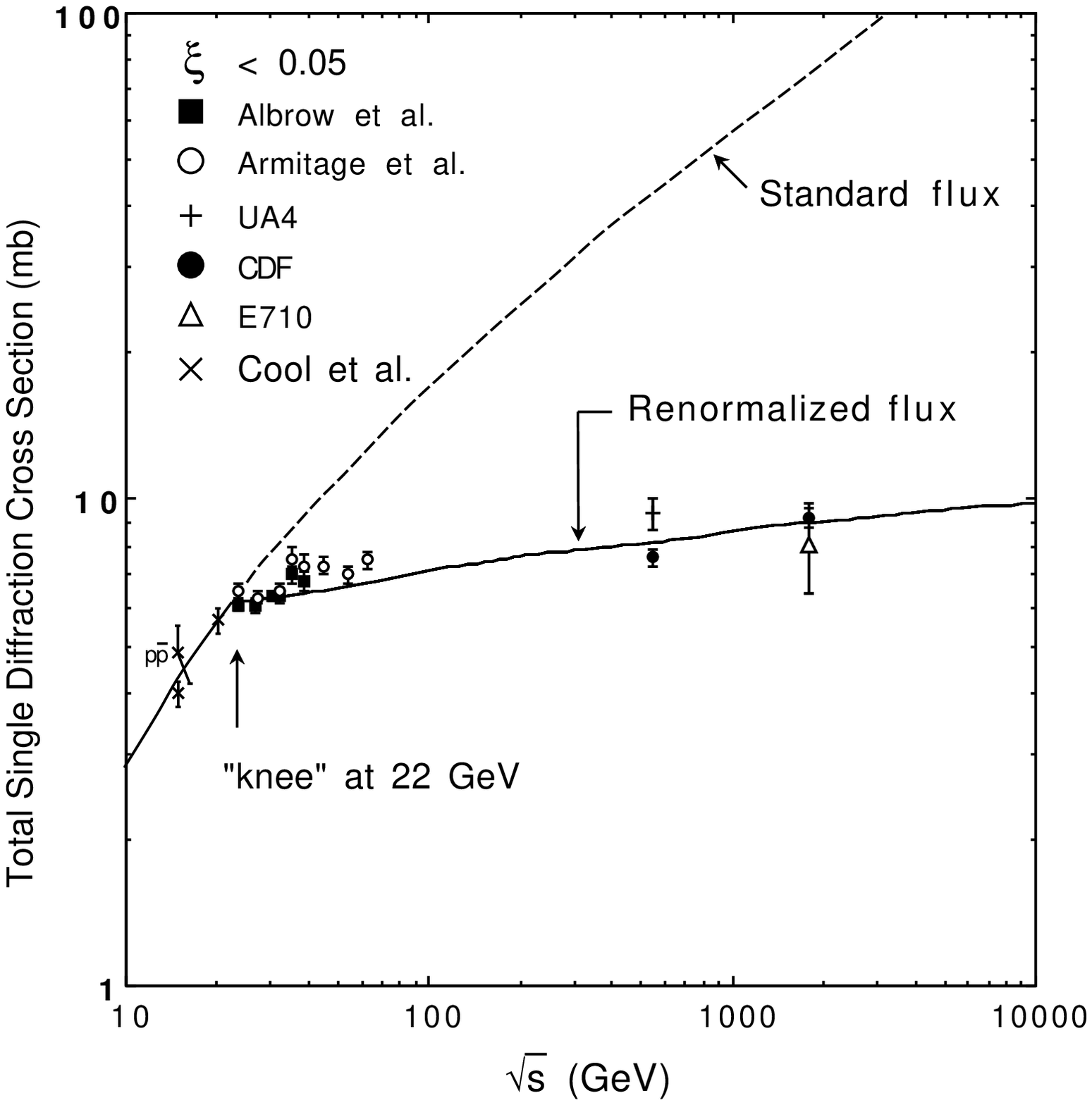,width=0.5\textwidth}}
\vspace*{-4em}
\caption{{\em (left)} Simultaneous fit to $p\bar p$, $\pi^\pm$, and $K^\pm$ total cross section and $\rho$-value data using eikonalized (solid) and Born level (dashed) amplitudes~\cite{ref:CMG} - the rise of the $p\bar p$ cross section with $\sqrt s$ is ``pulled'' by the rise of the $\pi^\pm$ cross sections and would pass through the CDF point at $\sqrt s=1800$ GeV even if this point were not used in the fit; {\em (right)} total \protect{$pp/p\bar p$} single diffraction dissociation cross section data (sum of $\bar p$ and $p$ dissociation) for \protect{$\xi<0.05$} compared with Regge predictions based on standard and renormalized Pomeron flux~\protect\cite{R}.}
\label{fig:sigmatot}
\end{figure}

Figure~\ref{fig:sigmatot}~({\em left}) shows Regge based fits to total and elastic scattering data using the eikonal approach to ensure unitarity~\cite{ref:CMG}. Good fits are obtained, which are consistent with the CDF cross sections at the Tevatron even if the Tevatron cross sections are not used in the fit~\cite{ref:CMG}. In contrast, the standard Regge fit to total single diffractive cross sections, shown in Fig.~\ref{fig:sigmatot}~({\em right}), overestimates the Tevatron cross sections by a factor of $\sim 10$. This discrepancy represents a breakdown of factorization, which is restored by the renormalization procedure proposed in Ref.~\cite{R} and discussed in Sec.~\ref{phenomenology}.   

\vspace*{-1em}
%
\subsection{Run Ia,b,c Results}
The diffractive processes studied by CDF in Tevatron Runs~Ia,b,c (1992-1996) are schematically shown in Fig.~\ref{fig:diagrams}b. Both soft and hard processes were studied. A discussion of the results obtained and of their significance in deciphering the QCD nature of the diffractive exchange can be found in Ref.~\cite{ref:lathuile}. The most interesting discoveries from this diffractive program were the breakdown of factorization and the restoration of factorization in events with multiple rapidity gaps.    

\paragraph{Breakdown of factorization.}
At $\sqrt s=$1800 GeV, the SD/ND ratios (gap fractions)   
for dijet, $W$, $b$-quark, and $J/\psi$ production, as well the ratio of
DD/ND dijet production, are all $\approx 1\%$.
This represents a suppression of a factor of $\sim$10 
relative to predictions based on 
diffractive parton densities measured from DDIS at HERA, indicating a breakdown of QCD factorization comparable to that
observed in soft diffraction processes relative to Regge theory expectations.  However, factorization approximately holds among the four different diffractive processes at fixed~\,$\sqrt s$, which indicates that the suppression  has to do with the formation of the rapidity gap, as predicted by the generalized gap renormalization model~(see~\cite{ref:lathuile} and Sec.~\ref{phenomenology}). 
\paragraph{Restoration of factorization in multi-gap diffraction.}
Another interesting aspect of the data is that 
ratios of two-gap to one-gap cross sections 
for both soft and hard processes obey factorization. This provides both a clue to understanding diffraction in terms of a composite Pomeron and 
an experimental  tool for diffractive studies using processes with 
multiple rapidity gaps~(see \cite{ref:lathuile}).

\subsection{The Run~II Diffractive Program} 
In Run~II, CDF has been conducting the following studies of diffraction:

$-$ structure function in dijet production,

$-$ $t$ distributions, 

$-$ exclusive dijet, $\gamma\gamma$, and $e^+e^-$ production,

$-$ structure function in $W$ production,

$-$ gap between jets: cross section vs. gap size for fixed $\Delta \eta^{jet}$. 

In this paper, we present preliminary results on the first three topics. The diffractive $W$ and `gap between jets' analyses are in progress and results are expected by Fall of 2007.

\subsection{Run II forward detectors}
\begin{figure}[htp]
\vglue -2.75 cm
{\hspace{-0.4cm}\psfig{figure=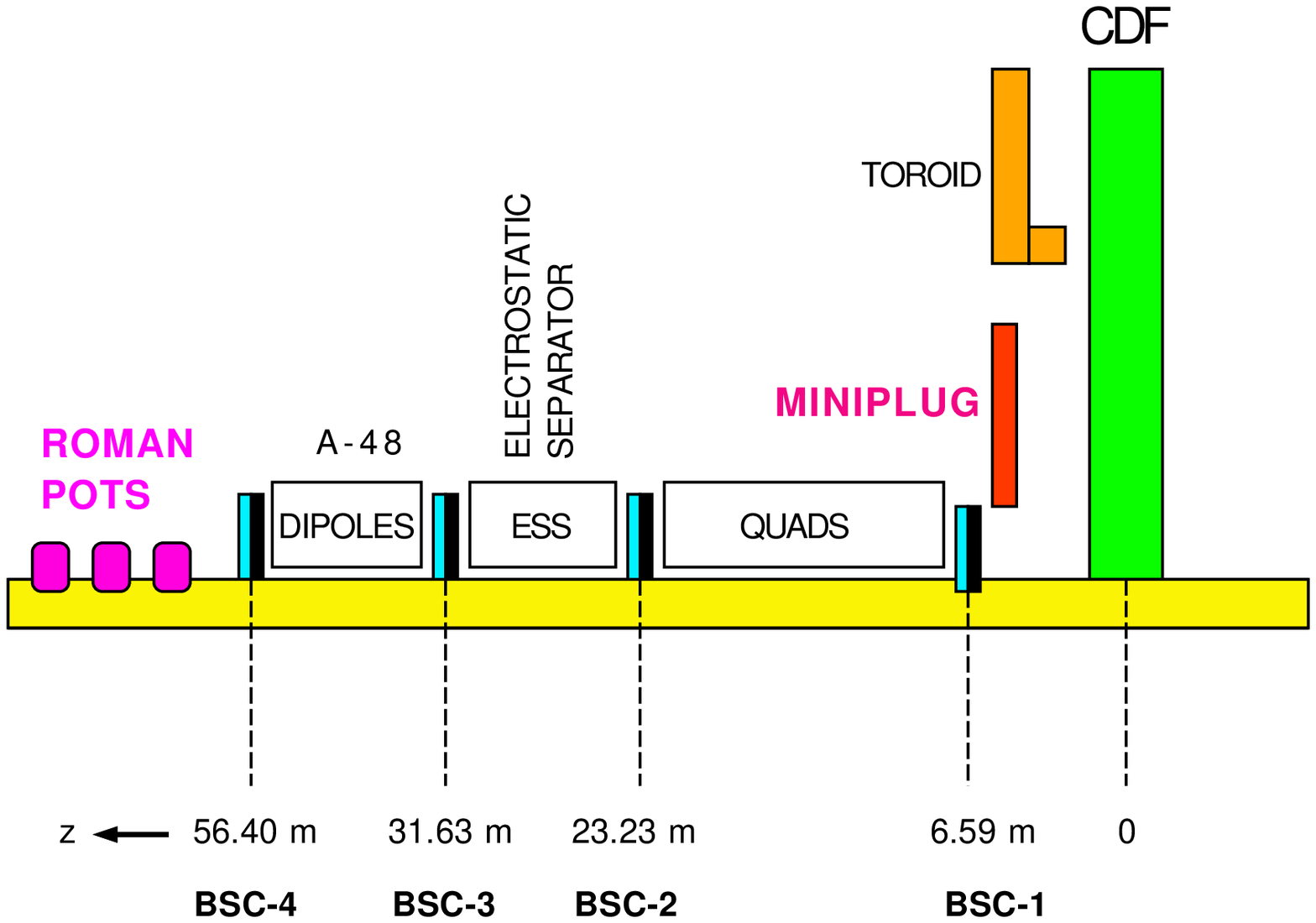,width=8.25cm}}
\vglue -25em
{\hspace{0.58\textwidth}\psfig{file=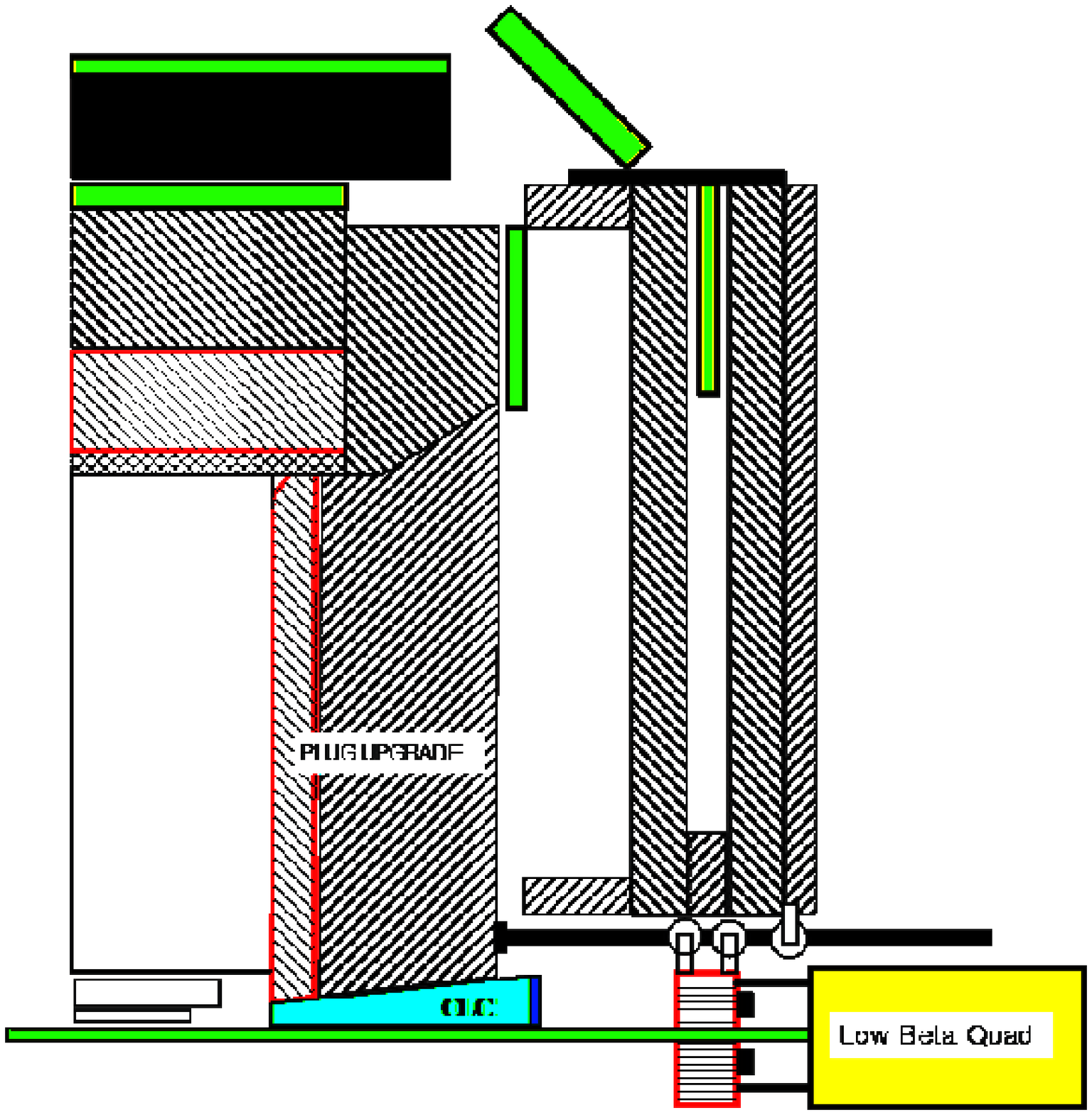,width=5cm}} 
\vglue -0.5em
{\hspace*{0.71\textwidth}CLC}{\hspace*{0.06\textwidth}MP}
\caption{The CDF detector in Run II: {\em (left)} location of forward detectors along the $\bar p$ direction; {\em (right)} position of the Cerenkov Luminosity Monitor (CLC) and MiniPlug calorimeters (MP) in the central detector.}
\label{BSC}
\end{figure}

The Run II diffractive program was made possible by an upgraded CDF detector~\cite{cdfrun2}, which includes the following special forward components (Fig.~\ref{BSC}):

$-$ Roman Pot Spectrometer (RPS) to detect leading antiprotons,

$-$ MiniPlug (MP) forward calorimeters approximately covering the region 
$3.5<|\eta|<5.5$,

$-$ Beam Shower Counters (BSC) positioned
around the beam pipe at four (three) locations along the $\bar p$ ($p$) 
beam direction to tag rapidity gaps within $5.5<|\eta|<7.5$.

\paragraph {The Roman Pot Spectrometer} 
is the same one that was used in Run Ic. It consists of 
$X$-$Y$ scintillation fiber detectors placed in three Roman Pot Stations
located at an average distance of $57$~m downstream in the $\bar p$ direction.
The detectors have a position resolution of $\pm 100\,\mu m$, which makes  
possible a $\sim 0.1\%$ measurement of the $\bar p$ momentum.    
In Run Ic, the $\bar p$-beam was behind the proton beam, as viewed
from the RPS side. Inverting the polarity (with respect to Run I) 
of the electrostatic beam separators enabled
moving the  RPS detectors closer to the $\bar p$-beam
and thereby obtain good acceptance for $|t|<0.5$ GeV$^2$  down to  
$\xi\equiv 1-x_F(\bar p)=0.03$ (for larger $|t|$, lower $\xi$ values can be reached).

\paragraph{The MiniPlug calorimeters}  are located within the holes of the muon toroids.
They consist of layers of lead plates immersed in 
liquid scintillator. The scintillation light is picked up by wavelength shifting fibers strung through holes in the lead plates and 
read out by multi-channel PMT's. 
The calorimeter ``tower" structure is defined by arranging fibers in groups 
to be read out by individual PMT pixels. There are 84 towers in each MiniPlug, and the signals they provide can be used to measure energy and position for both electromagnetic and hadron initiated showers    
~\cite{MP}.
 
\paragraph{The Beam Shower counters} are rings of scintillation
counters ``hugging" the beam pipe.  The BSC-1 rings are 
segmented into four quadrants, while all other BSC rings are segmented into two halves.
The BSC-1 are also used to provide rapidity gap triggers and for measuring beam losses.  

\subsection{Diffractive structure function from dijet production}
\begin{center}
\begin{figure}[hp]
\vglue -1em
\hbox{\psfig{figure=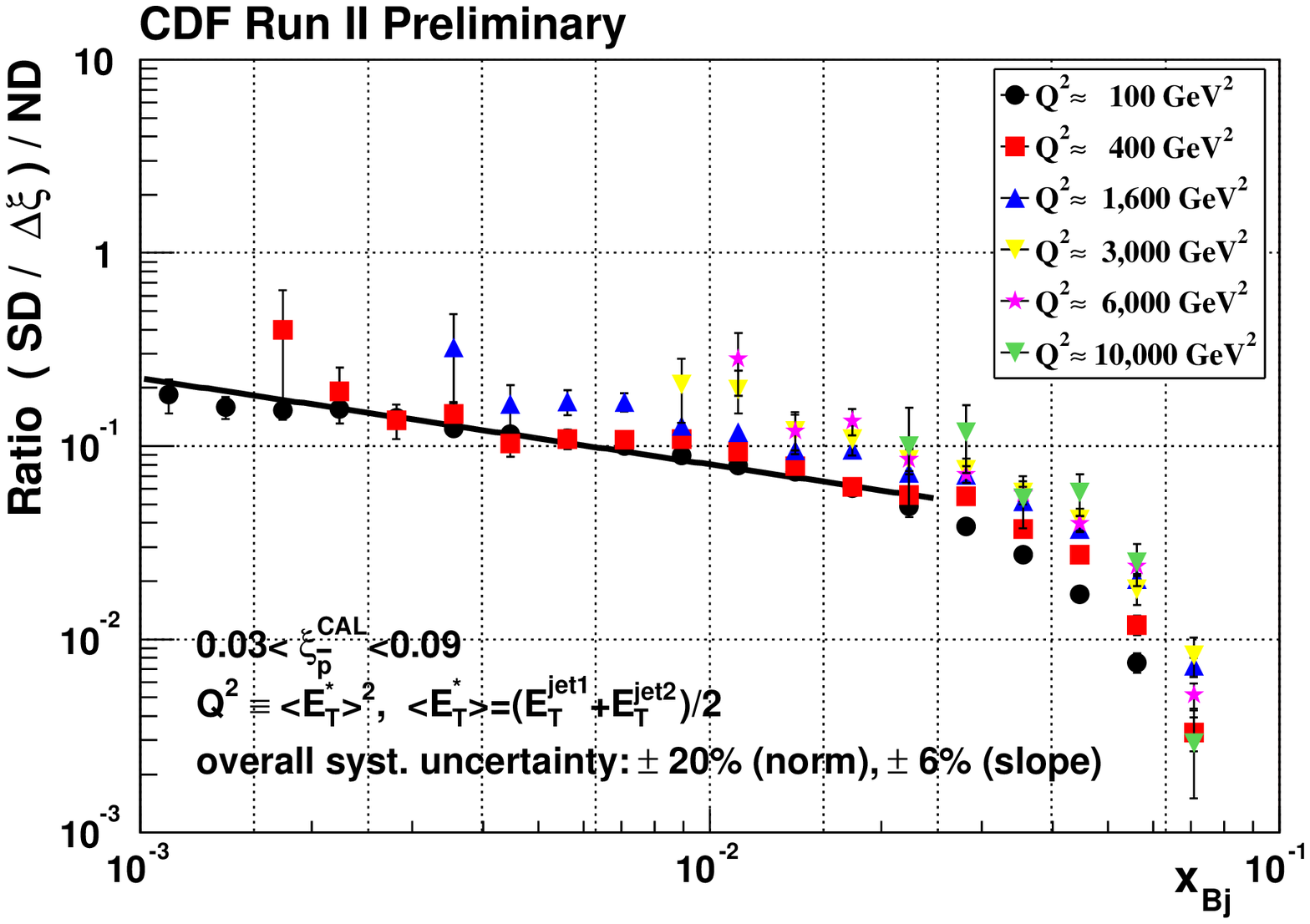,width=0.53\textwidth}\psfig{figure=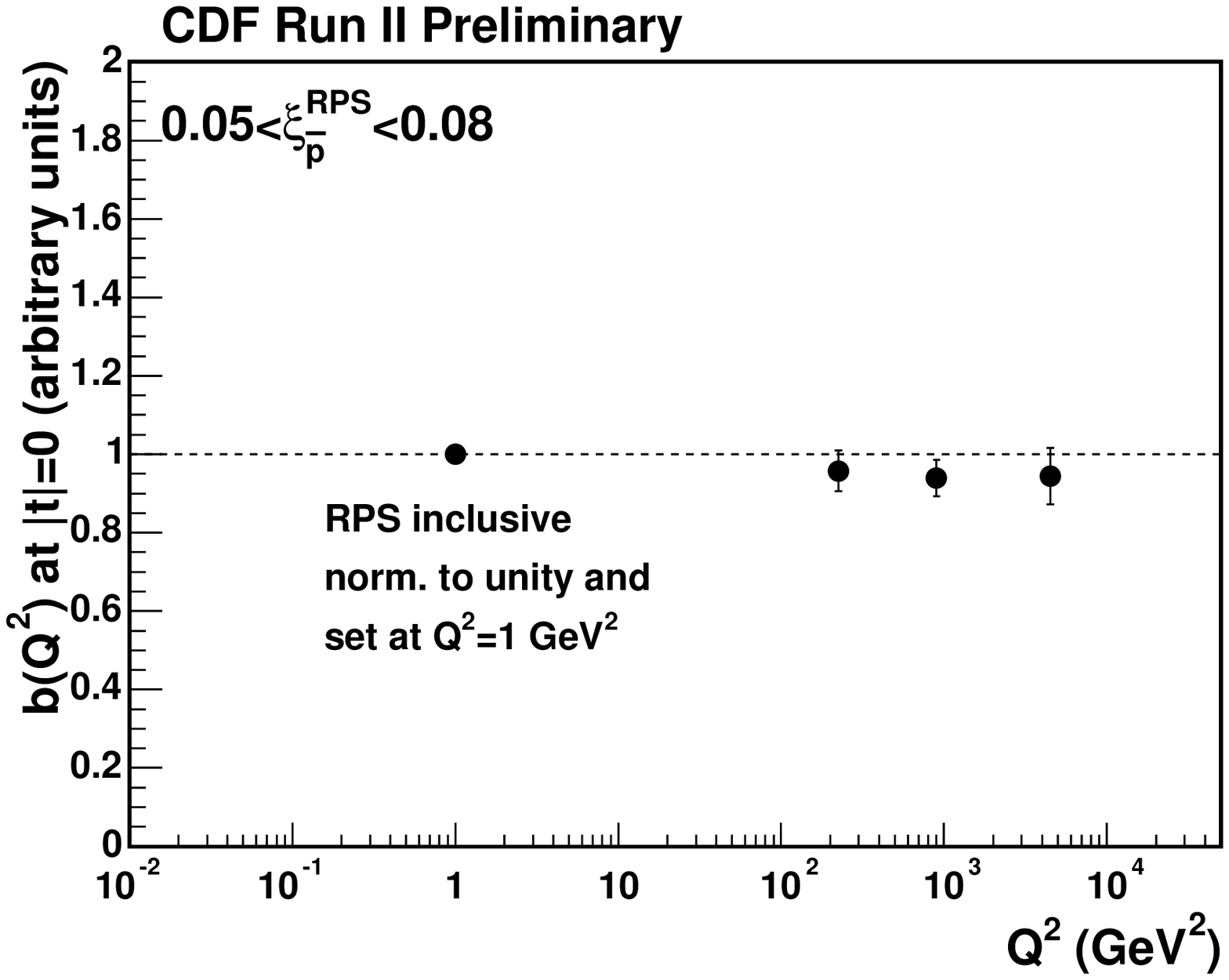,width=0.5\textwidth}}
\caption{
{\em (left)} Ratio of diffractive to non-diffractive dijet event rates as a function of $x_{Bj}$ (momentum fraction of parton in antiproton) for different values of $E_T^2=Q^2$;
{\em (right)} the $b(t)|_{t=0}$ slope vs. $Q^2$. 
}
\label{fig:xbjQ2}
\end{figure}
\end{center}

In Run II, CDF has obtained preliminary results for the $x_{Bj}$, $Q^2$, and $t$ dependence of the diffractive structure function from dijet production at $\sqrt s=1960$ GeV.  The measured $x_{Bj}$ rates confirm the factorization breakdown observed in Run~I (see review in Ref.~\cite{ref:michgallDIS}). 
The $Q^2$ and $t$ dependence results are shown in Fig~\ref{fig:xbjQ2}. 
\vspace*{-3em}
\paragraph{$Q^2$ dependence.} In the range  $10^2\hbox{ GeV}^2<Q^2<10^4$~GeV$^2$, where the inclusive $E_T$ distribution falls by a factor of $\sim 10^4$, the ratio of the SD/ND distribution increases, but only by a factor of $\sim 2$. This result indicates that the $Q^2$ evolution in diffractive interactions is similar to that in ND interactions. 
\paragraph{$t$-dependence.} The slope parameter $b(Q^2,t)|_{t=0}$ of an exponential fit to $t$ distributions near $t=0$ shows no $Q^2$ dependence in the range $1\hbox{ GeV}^2<Q^2<10^4\hbox{ GeV}^2$.

 These results support the picture of a composite Pomeron formed from color singlet combinations of the underlying parton densities of the nucleon (see~\cite{ref:lathuile} and Sec.~\ref{phenomenology}).        

\subsection{Exclusive Dijet Production}
\begin{figure}[htp]
\centerline{\hbox{\psfig{figure=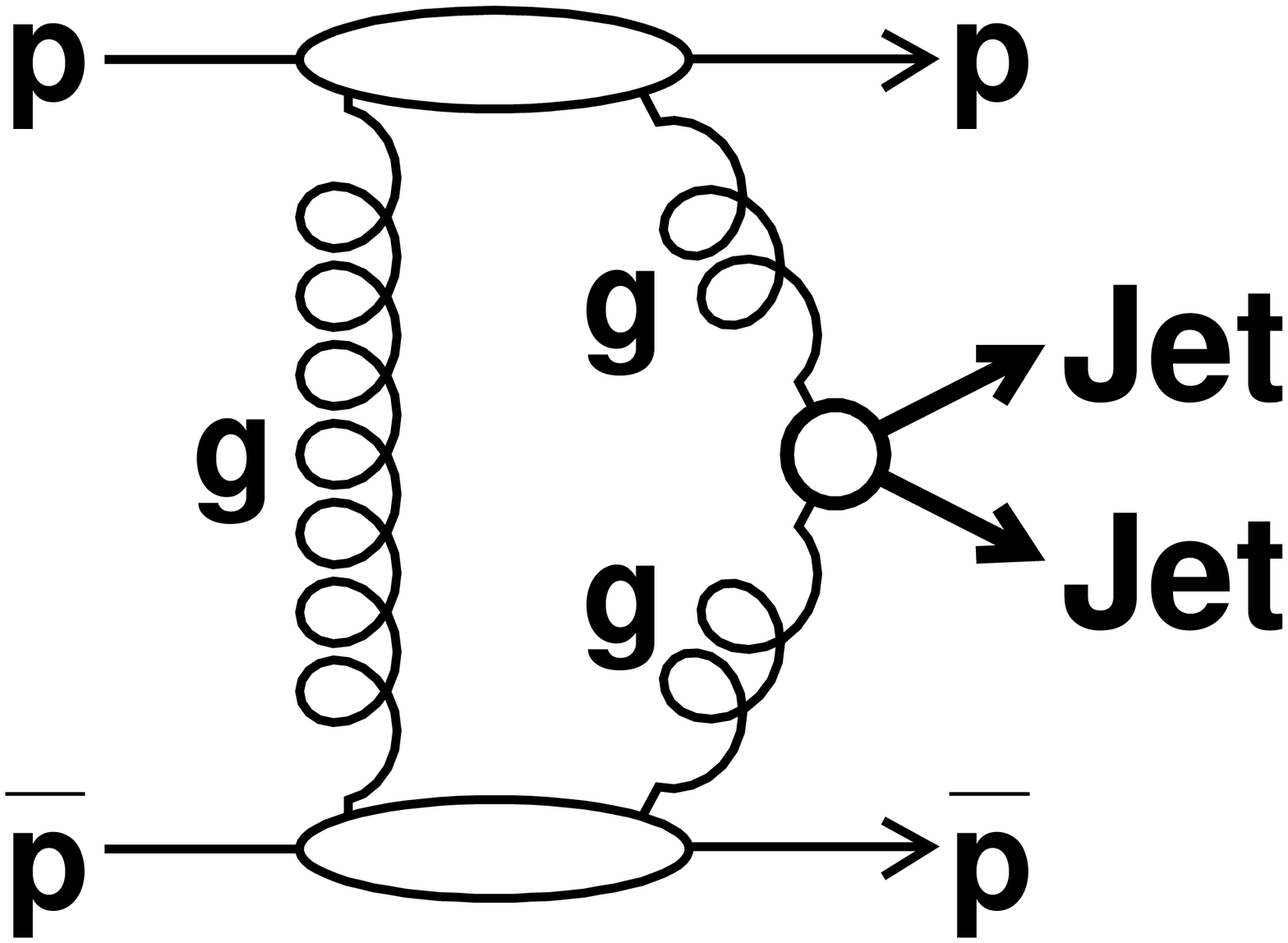,width=0.25\textwidth}\hspace*{2cm}\psfig{figure=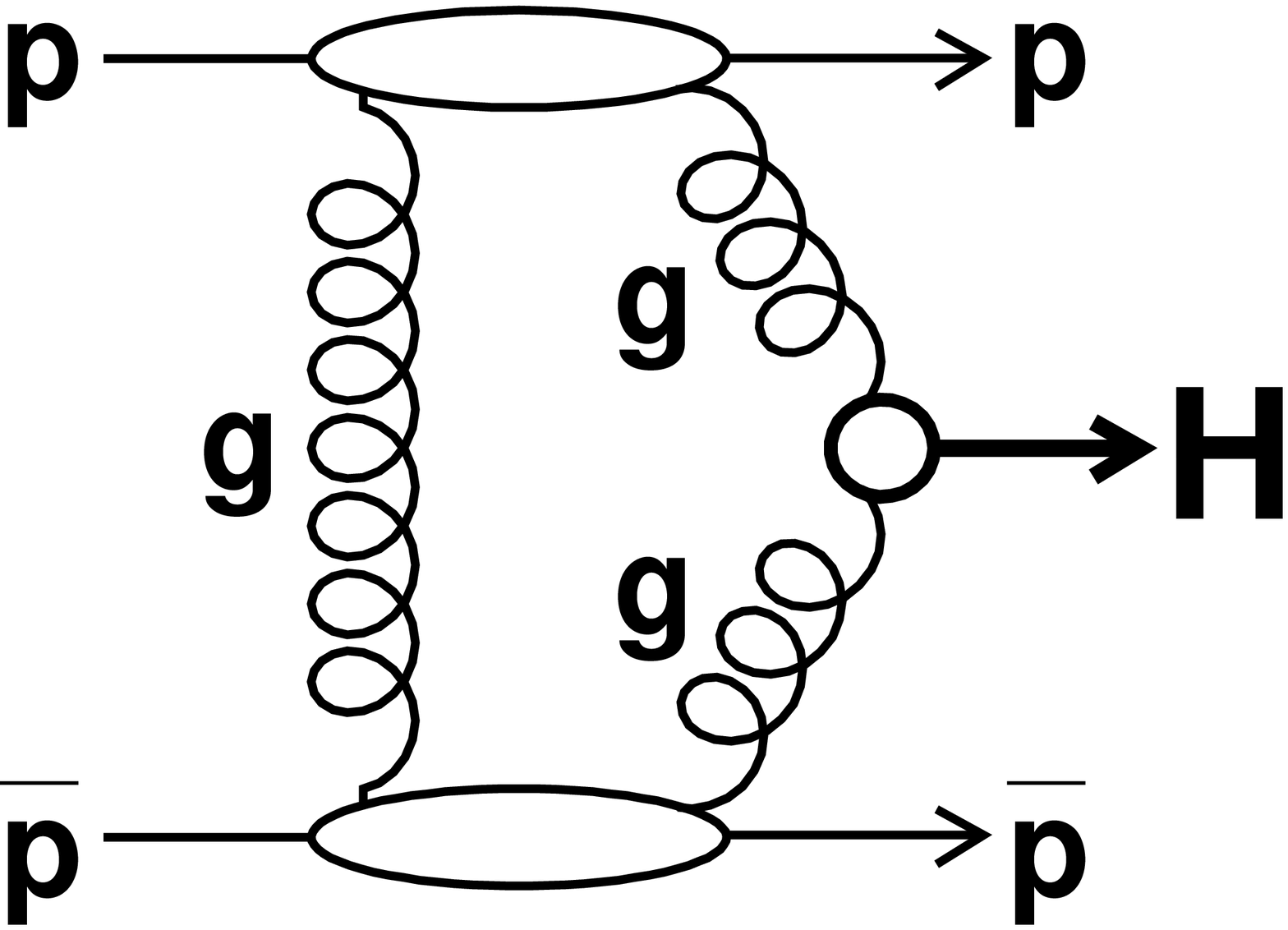,width=0.25\textwidth}}}
\caption{Lowest order diagrams for exclusive dijet (left) and Higgs (right) production in $\bar{p}p$ collisions.} 
\label{fig:excl_diagram}
\end{figure}

Exclusive production in $p\bar p$ collisions is of interest not only for testing QCD inspired models of diffraction, but also as a tool for discovering new physics. The process that has attracted the most attention is exclusive Higgs boson production. The search for Higgs bosons is among the top priorities in the research plans of the LHC experiments. While the main effort is directed toward searches for inclusively produced Higgs bosons, an intense interest has developed toward  exclusive Higgs production, $\bar p/p+p\rightarrow \bar p/p+H+p$. This Higgs production channel presents the advantage that it can provide clean events in an environment of suppressed QCD background, in which the Higgs mass can accurately be measured using the missing mass technique by detecting and measuring the momentum of the outgoing proton and (anti)proton. 
However, exclusive production is hampered by expected low production rates~\cite{ref:KMR}. As rate calculations are model dependent and generally involve non-perturbative suppression factor(s), it is considered prudent to calibrate them  against processes involving the same suppression factors(s), but have higher production rates that can be measured at the Tevatron. One such processes is exclusive dijet production, which proceeds through the same mechanism as Higgs production, as shown in Fig.~\ref{fig:excl_diagram}.  

The search for exclusive dijets is based on measuring the 
dijet mass fraction, $R_{jj}$, 
defined as the mass of the two leading jets in an event, $M_{jj}$,  
divided by the total mass 
reconstructed from the energy deposited in all calorimeter towers, $M_X$.
The signal from exclusive diets
is expected to appear at high values of $R_{jj}$, smeared by resolution and radiation effects.  Events from 
inclusive DPE production, 
$\bar p p\rightarrow \bar p+gap+jj+X+gap$, are expected to contribute to the 
entire $M_{jj}$ region. Any such events within the exclusive $M_{jj}$ range contribute to background and must be subtracted when evaluating exclusive production rates.

The method used to extract the exclusive signal from the inclusive $R_{jj}$ distribution is based on fitting the data with MC simulations~\cite{MCprograms}. Two methods have been used. In the first one, the POMWIG and ExHuME generators are used for simulating inclusive and exclusive events, respectively; in the second, inclusive (exclusive) distributions are simulated using the POMWIG (DPEMC) program. Experimentally, the MC non-exclusive dijet background shape is checked by a study of high $E_T$ $b$-tagged dijet events, as quark jet production through $gg\rightarrow \bar qq$  is suppressed in LO and NLO QCD by the $J_z=0$ selection rule as $m_q/M^{jet}\rightarrow 0$. 
 
\begin{figure}[htp]
\centerline{
\hbox{
\psfig{figure=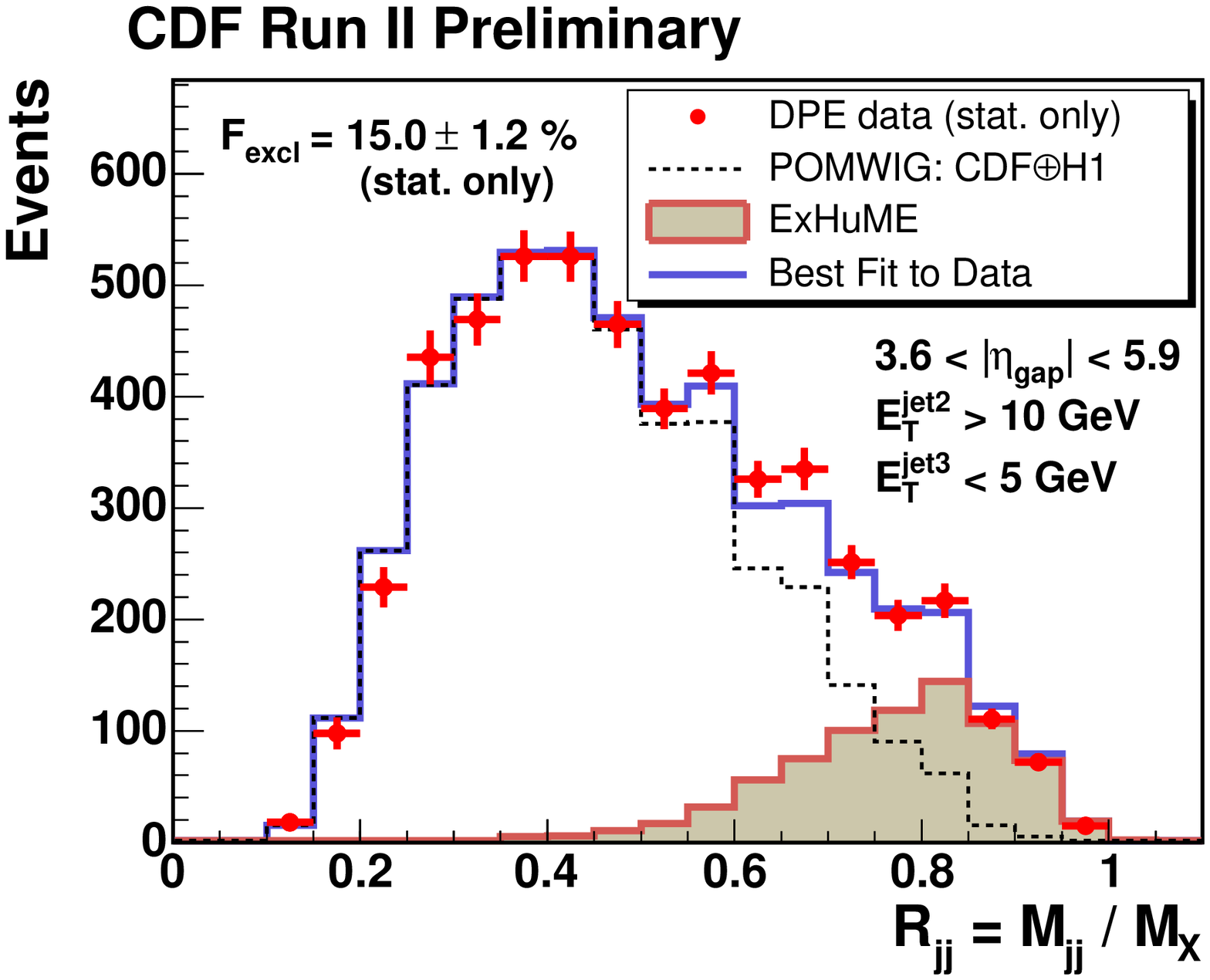,width=0.5\textwidth}
\hspace*{2em}\psfig{figure=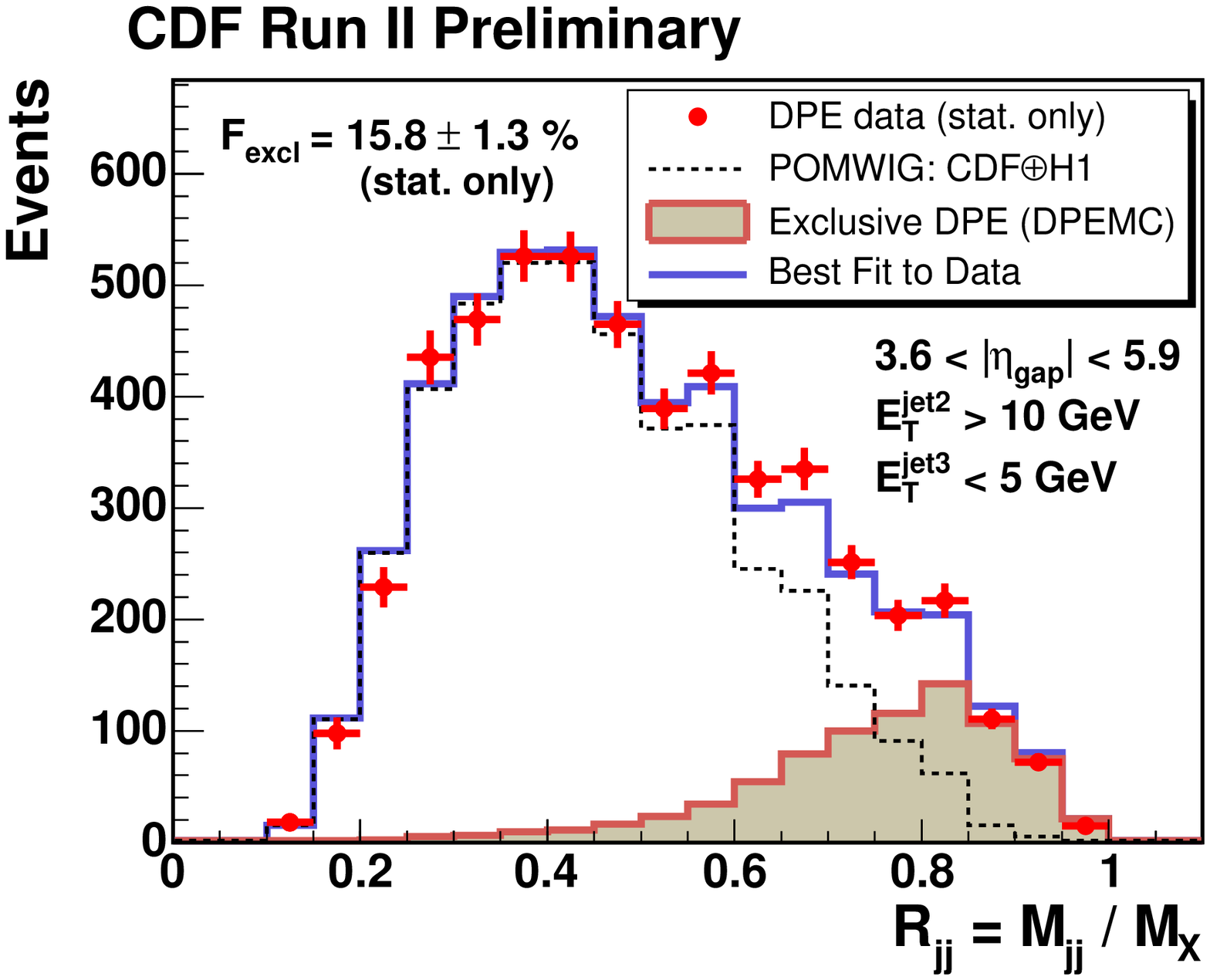,width=0.5\textwidth}
}}
\caption{Extraction of exclusive dijet production signal using Monte Carlo techniques to subtract the inclusive dijet background: {\em (left)} dijet mass fraction in data (points) and best fit (solid line) obtained from MC events generated using the POMWIG (dashed) and  ExHuME (filled) MC generators for inclusive and exclusive events, respectively; {\em (right)} the same data fitted with POMWIG and exclusive DPEMC generators.}
\label{fig:exclJJsignal}
\end{figure}

Figure~\ref{fig:exclJJsignal} shows measured $R_{jj}$ distributions plotted vs. dijet mass fraction. On the left, the number of events within the specified kinematic region is compared with fits based on POMWIG$\oplus$ExHuME distribution shapes, and on the right with fits based on POMWIG$\oplus$DPEMC predictions. Both approaches yield good fits to the data. The suppression factor expected for exclusive $b$-tagged dijet events is checked with CDF data in Fig.~7.  Within the quoted errors, this result validates the MC based method for extracting the exclusive signal.

\begin{figure}[htp]
\centerline{\psfig{figure=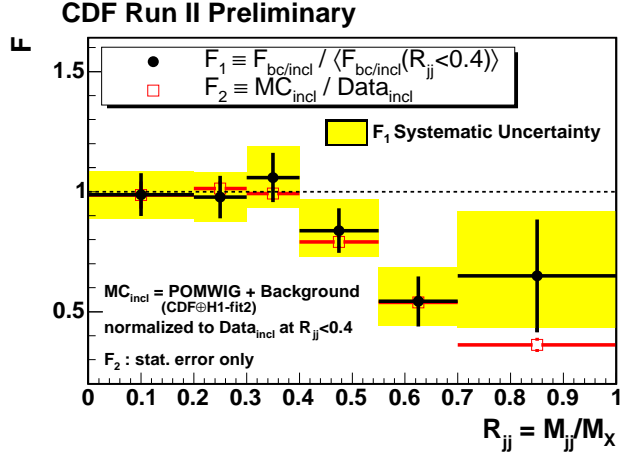,width=0.75\textwidth}}
\caption{{\em (circles)} Fraction of heavy flavor ($b,c$) in all dijet events in data, $F_1$, as a function of dijet mass fraction showing the expected suppression at high $M_{jj}$; {\em (squares)} fraction, $F_2$,  of inclusive MC in data from Fig.~\protect\ref{fig:exclJJsignal} (left). The agreement between the measured suppression levels in $F_1$ and $F_2$ serves to validate  the MC based technique of extracting the exclusive production rate from the data.}
\end{figure}
%
\begin{figure}[htp]
\centerline{
\hbox{
\psfig{figure=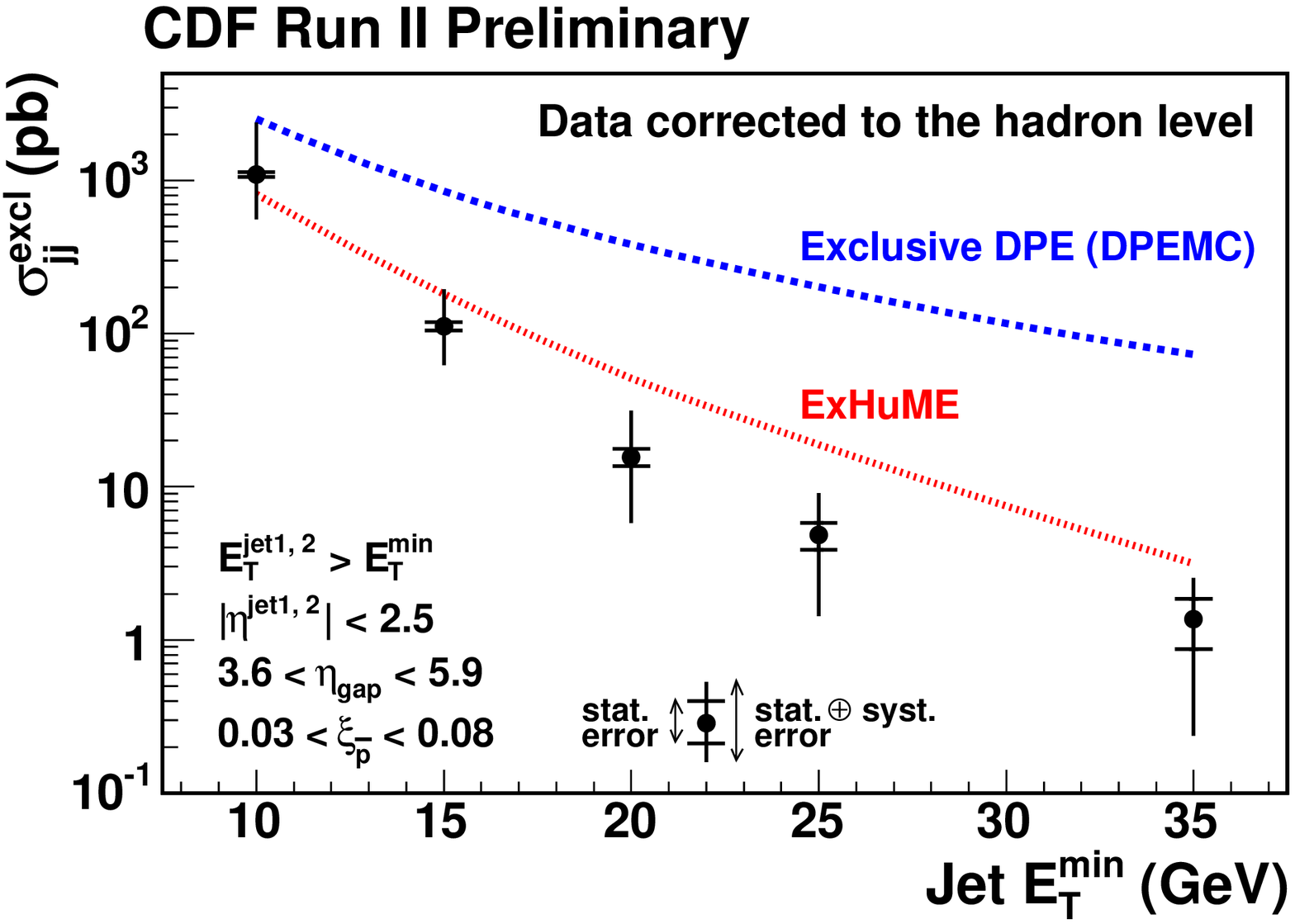,width=0.5\textwidth}
{\psfig{figure=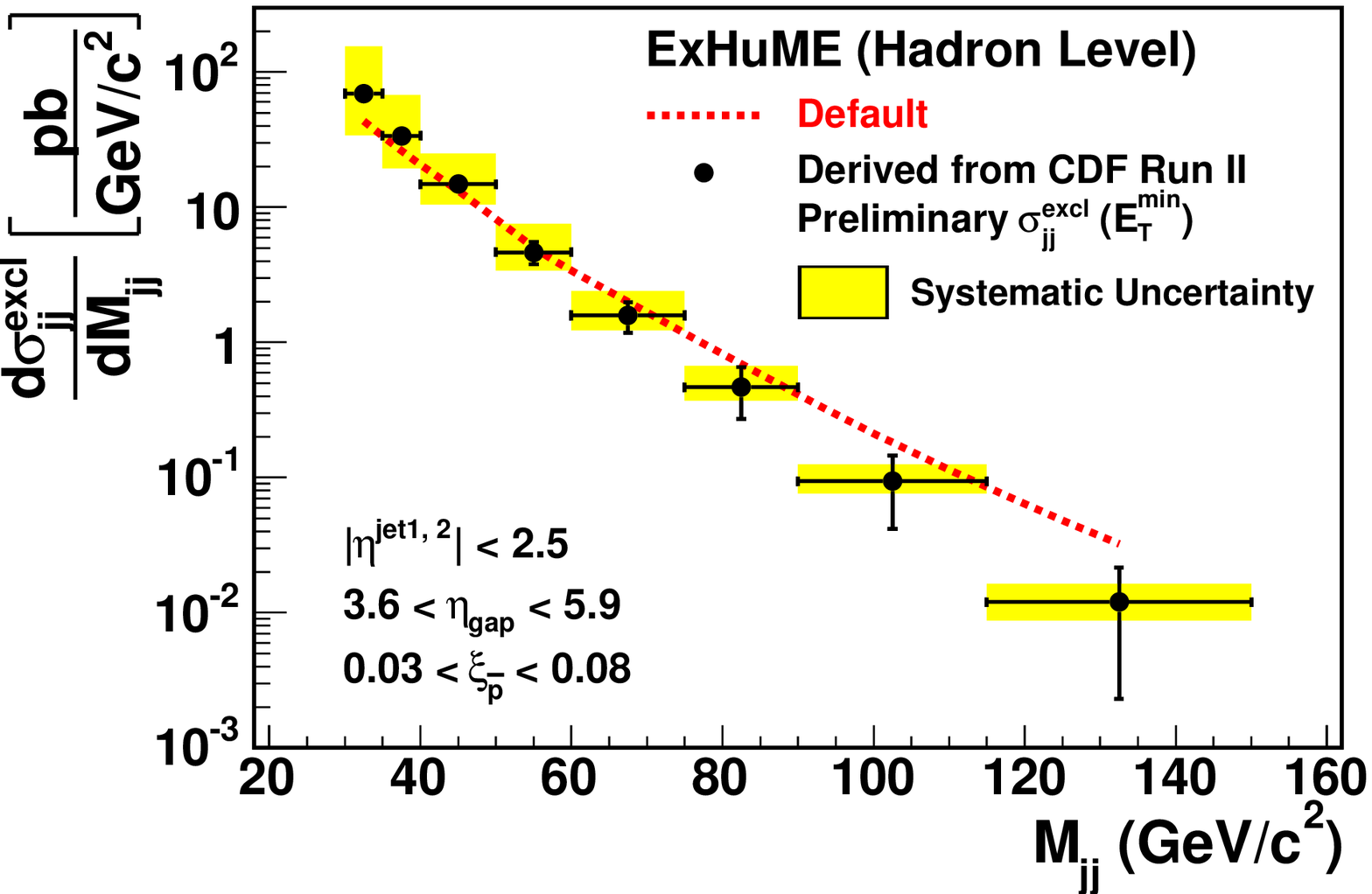,width=0.5\textwidth}}
}}
\caption{{\em (left)} Measured exclusive dijet cross sections vs. the minimum $E_T$ of the two leading jets compared with ExHuME and DPEMC predictions; {\em (right)} ExHuME hadron level differential exclusive dijet cross section vsersus dijet mass normalized to the CDF cross sections at {\em left}. The systematic errors shown are propagated from those in the data; the ExHuME predictions have comparable systematic uncertainties.}
\label{fig:ET_MJJ}
\end{figure}

In Fig.~\ref{fig:ET_MJJ}~{\em (left)}, integrated cross sections above a minimum $E_T^{jet1,2}$ are compared with ExHuME and DPEMC predictions. The data favor the ExHuME prediction. ExHuME hadron level differential cross sections $d\sigma^{excl}/dM_{jj}$ normalized to the measured data points of Fig.~\ref{fig:ET_MJJ}~{\em (left)} are shown in Fig.~\ref{fig:ET_MJJ}~{\em (right)} with errors propagated from the uncertainties in the data. Within the errors, the good agreement with the default ExHuME prediction up to masses in the region of the  standard model Higgs mass predicted from global fits to electroweak data lends credence to the calculation of Ref.~\cite{ref:KMR} for exclusive Higgs boson production at the LHC.

\subsection{Exclusive $\gamma\gamma$ and $e^+e^-$ production} 
Exclusive $\gamma\gamma$ production in $p\bar p$ collisions proceeds through a lowest order diagram similar to that of Fig.~\ref{fig:excl_diagram}~{\em (right)}, but with the gluons that produce the Higgs replaced by $\gamma$'s. Therefore, like exclusive dijet production, exclusive $\gamma\gamma$ production can also be used for calibrating models of Higgs production at hadron colliders. Exclusive $e^+e^-$ production is a QED process whose cross section can be reliably calculated and thus can serve validate the procedure used to extract the exclusive $\gamma\gamma$ signal. 

A search for exclusive $\gamma\gamma$ production has been performed on a sample of events collected by requiring a high $E_T$ electromagnetic shower in combination with a loose forward rapidity gap requirement. In the data analysis, the rapidity gap requirement was tightened, and the search was narrowed down to  events with two high $E_T$ photon showers satisfying certain ``exclusivity'' requirements. In a data sample of 532 pb$^{-1}$ total integrated luminosity, three exclusive $\gamma\gamma$ candidate events with $E^\gamma_T>5$~GeV were found with no tracks pointing to the electromagnetic clusters. As a check of the robustness of the rapidity gap requirement, CDF measured the cross section for the  purely QED process $\bar p+p\rightarrow \bar p+e^+e^-+p$~\cite{excl_ee}. Twelve exclusive $e^+e^-$ candidate events were found in the data with an estimated background of $2.1^{+0.7}_{-0.3}$, yielding $\sigma (e^+ e^-)= 1.6^{+0.5}_{-0.3} {\rm (stat)}$, which agrees with an expectation of $1.71\pm 0.01$~pb. For $\gamma\gamma$ production, three candidate exclusive events were observed, yielding an upper limit on the production cross section of 110~fb at 95\% confidence level.

\subsection{Summary of experimental results section}
The diffractive program of the CDF Collaboration at the Fermilab Tevatron $p\bar p$ Collider has been reviewed with emphasis on recent results from Tevatron Run~II at $\sqrt s=$1.96~TeV. 

Run~I results have been briefly presented and their physics significance placed in perspective. Processes studied by CDF in Run~I include elastic and total cross sections, soft diffractive cross sections with single and multiple rapidity gaps, and hard single diffractive production of dijet, $W$, $b$-quark, and $J/\psi$ production, as well as central dijet production in events with two forward rapidity gaps (double Pomeron exchange). The results obtained support a picture of universality of diffractive rapidity gap formation across soft and hard diffractive processes, which favors a composite over a particle-like Pomeron made up from color singlet quark and/or gluon combinations with vacuum quantum numbers.    
 
Run~II results on the $x_{Bj}$ and $Q^2$ dependence of the diffractive structure function obtained from dijet production have been presented, as well as on the slope parameter of the $t$-distribution  of diffractive events as a function of $Q^2$. In the range  $10^2\hbox{ GeV}^2<Q^2<10^4$~GeV$^2$, where the inclusive $E_T$ distribution falls by a factor of $\sim 10^4$, the ratio of SD/ND distributions varies by only a factor of $\sim2$, indicating that the $Q^2$ evolution in diffractive interactions is similar to that in ND ones; and the slope parameter $b(Q^2)|_{t=0}$ of an exponential fit to $t$ distributions near $t=0$ in the range $1\hbox{ GeV}^2<Q^2<10^4\hbox{ GeV}^2$ shows no $Q^2$ dependence. These results support a picture of a composite diffractive exchange (Pomeron) made up from the underlying parton densities of the nucleon.

Results on cross sections for exclusive dijet, $\gamma\gamma$, and $e^+e^-$ production have  also been presented and their significance for calibrating theoretical estimates for exclusive Higgs production at the Large Hadron Collider discussed. The exclusive dijet cross section was measured up to jet $E_T^{min}$ of 35 GeV. When expressed as a function of dijet mass $M_{jj}$, cross sections up to masses of $M_{jj}\sim 120-140$~GeV are obtained, which are in the region of the standard model Higgs mass expected from global fits to electroweak data. A measurement of exclusive $\gamma\gamma$ production, a process which can also be used for calibrating Higgs production models, yielded three exclusive candidate events, setting an upper limit on the production cross section in the neighborhood  of that predicted in Ref.~\cite{ref:KMR}.

\section{Phenomenology\label{phenomenology}}
\subsection{Introduction\label{intro}}
As mentioned in Sec.~\ref{experiment}, adronic diffraction is characterized by one or more large rapidity gaps, defined as regions of (pseudo)rapidity
devoid of particles.
 Rapidity gaps may occur in non-diffractive (ND) interactions by fluctuations in particle multiplicity. However, the probability for such occurrences is expected to be exponentially suppressed as a function of gap width, since at a given $\sqrt s$ the particle density $\rho_N=dN/d\eta$ is approximately constant vs. $\eta$ and therefore the probability for no particles being produced at ($\eta,\Delta\eta$) is  by Poisson statistics $P_0(\eta,\Delta\eta)=e^{-{\rho_N\Delta\eta}}$~\cite{Bj}. 
Diffractive rapidity gaps do not exhibit such a suppression. This aspect of diffraction is attributed to the lack of radiation in the diffractive exchange, which proceeds through a colorless quark/gluon construct with vacuum quantum numbers, historically referred to as the {\em Pomeron}~\cite{books}. In this section, we use the term {\em Pomeron} generically to denote a diffractive exchange, and describe it in terms of the underlying parton densities of the proton as a function of $Q\;^2$. Our  phenomenological description is driven by regularities observed in results on soft and hard diffractive processes obtained in $pp\,/\,p\bar p$ collisions in fixed target and collider experiments, and in $ep$ collisions at HERA (see review in Ref.~\cite{ref:lathuile}).
\begin{figure}[h]
\psfig{figure=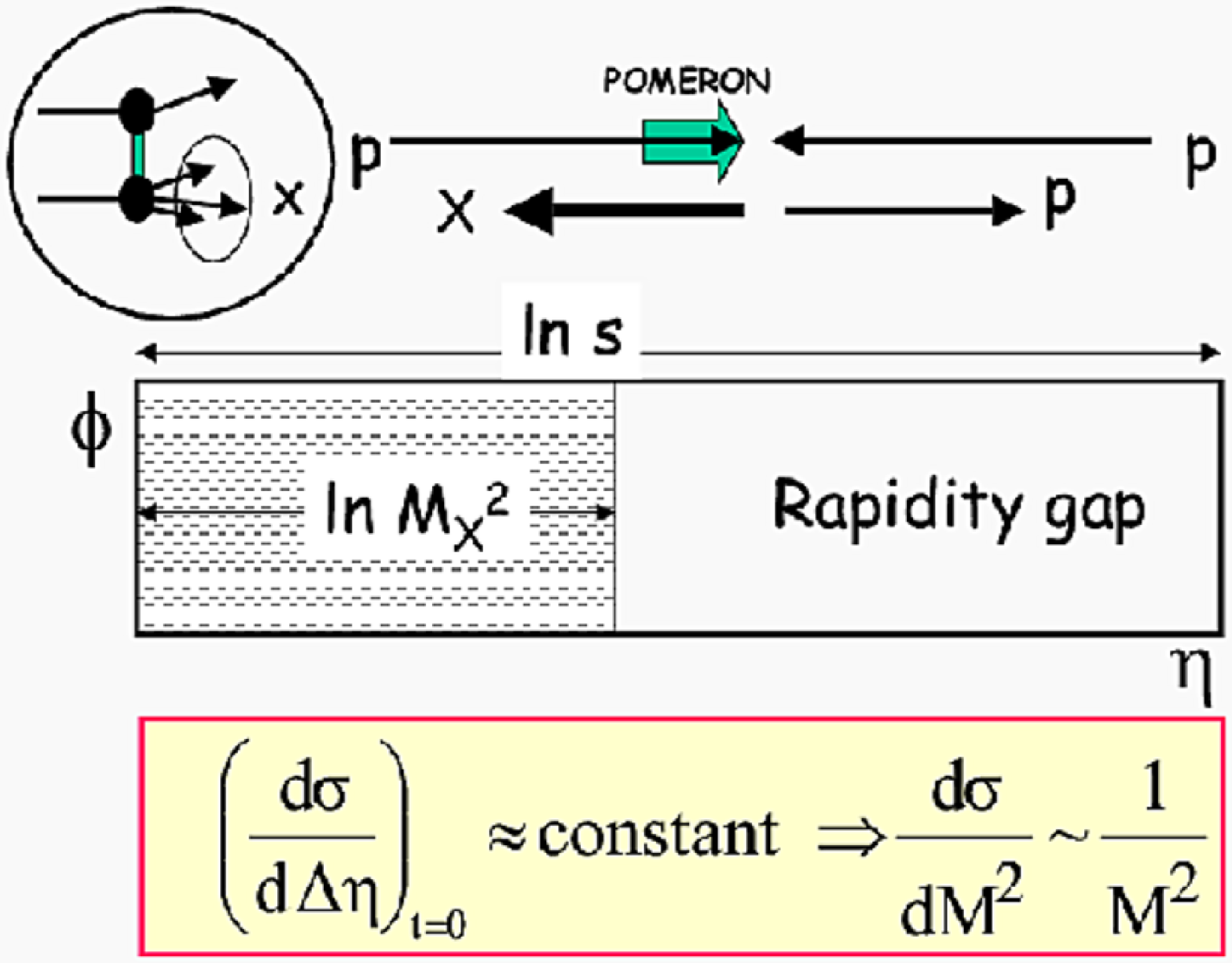,width=0.5\textwidth}
\vspace*{-5cm}
{\hspace*{6cm}\psfig{figure=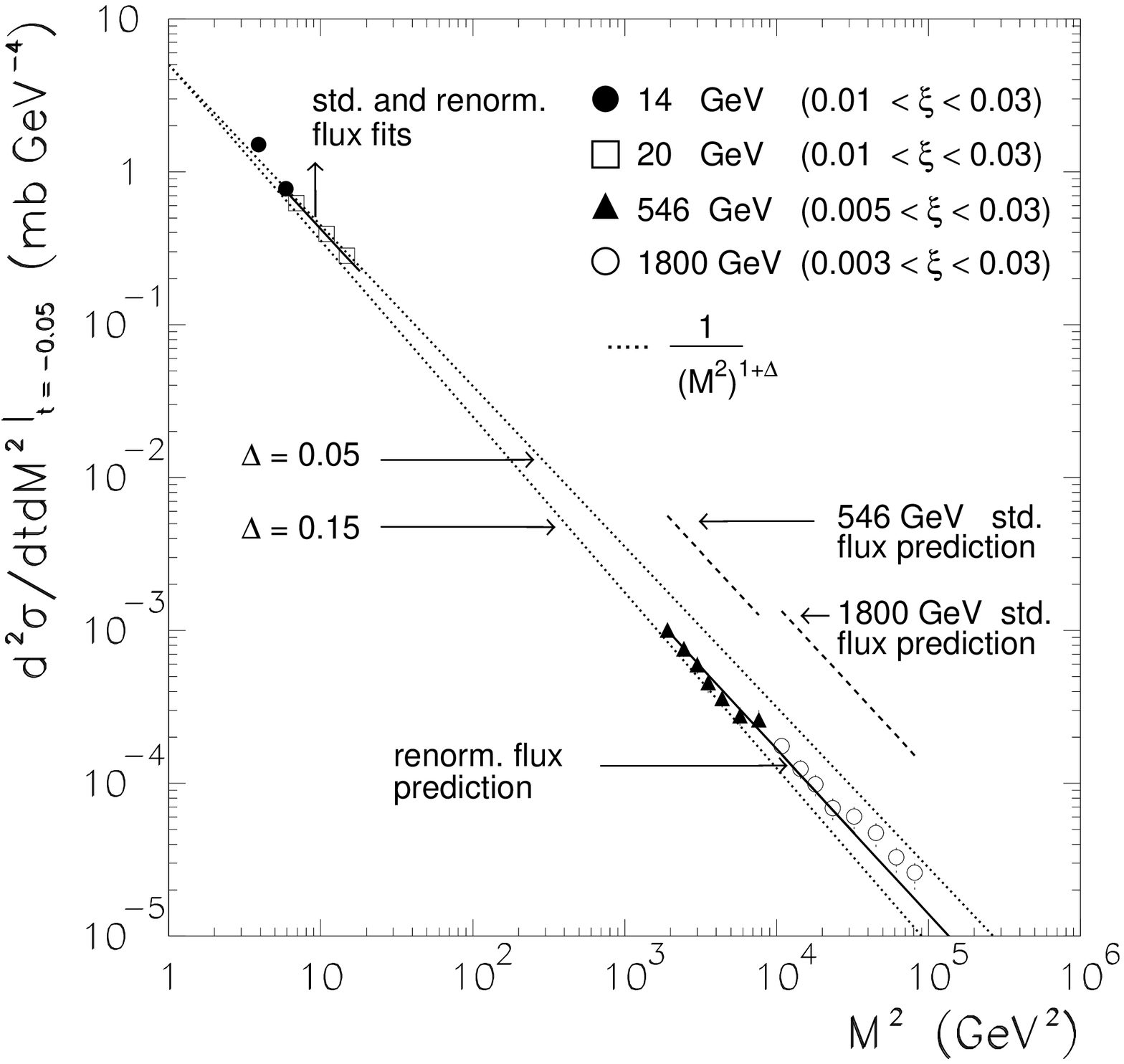,width=0.5\textwidth}}
\vspace*{-2em}
\caption{{\em (left)} Event topology for $pp$ single diffraction dissociation, $pp\rightarrow pX$: lack of radiation in ``vacuum exchange'' leads to a cross section independent of $\Delta \eta$ and thereby to  $1/M^2$ behavior;  {\em (right)}~Cross sections \protect$d^2\sigma_{sd}/dM^2 dt$ 
for $p+p(\bar p) \rightarrow p(\bar p)+X$ at
$t=-0.05$ GeV$^2$ and $\protect\sqrt s=14$, 20, 546 and 1800 GeV 
compared with the renormalized Pomeron flux prediction\cite{GM} ($\Delta$ here is the Pomeron intercept, denoted by $\epsilon$ through the rest of this report). 
At $\protect\sqrt s$=14 and 20 GeV,
the fits using the standard and renormalized fluxes coincide; 
at the higher energies, the standard flux prediction 
overestimates the cross sections by~$\sim {\cal{O}}(10)$.}
\label{fig:rapgap}
\end{figure}
A clue to understanding diffraction in QCD is provided by the $d\sigma/dM^{\;2}$ behavior of the soft single diffractive cross section. As shown in Fig.~\ref{fig:rapgap}~{\em (left)}, due to absence of radiation in vacuum exchange, a $d\sigma/dM^2\sim 1/M^2$ behavior is expected with no explicit $s$-dependence. This is observed in the data plotted in ~Fig.~\ref{fig:rapgap}~{\em (right)}. The deviation from exact $1/M^{\;2}$ behavior holds the key to understanding diffractive cross sections in terms of the underlying parton densities of the diffracted nucleon.         

\subsection{The data}  
Scaling and factorization properties observed in data provide the input to our model of describing hadronic diffraction in terms of inclusive parton densities and QCD color factors. The foundations of the model are mainly results from the CDF experiment at the Tevatron $p\bar p$ collider at Fermilab, and from the H1 and ZEUS experiments at the HERA $ep$ collider at DESY. 

Table~1 lists the soft diffractive processes studied by CDF in Run~I. Measurement details and results can be found in the references provided.
\begin{table}[h]\label{tab:soft}
\begin{center}
\caption{Soft diffractive processes studied by CDF }
\vspace*{0.5em}
\begin{tabular}{lll}
{\bf ND}&Non-Diffractive $(\sigma_T)$~\cite{tot}&$\stackrel{-}{p}+p\rightarrow X\,^\dagger$\\
{\bf EL}&Elastic~\cite{el}&$\stackrel{-}{p}+p\rightarrow \stackrel{-}{p}+p$\\
{\bf SD}&Single Diffraction~\cite{sd}&$\stackrel{-}{p}+p\rightarrow \stackrel{-}{p}+{\rm gap}+X$\\
{\bf DD}&Double Diffraction~\cite{dd}&$\stackrel{-}{p}+p\rightarrow X+{\rm gap}+Y$\\
{\bf DPE}&Double Pomeron Exchange~\cite{idpe}&$\stackrel{-}{p}+p\rightarrow \stackrel{-}{p}+{\rm gap}+X+{\rm gap}+p$\\
{\bf SDD}&Single plus Double Diffraction~\cite{sdd}&
$\stackrel{-}{p}+p\rightarrow \stackrel{-}{p}+{\rm gap}+X+{\rm gap}+Y$\\
\end{tabular}
\vglue 0.2cm
\centerline{$^\dagger$ $\sigma_T$ is included since by the optical theorem it is related to $\hbox{Im  }f^{\;el}(t=0)$}
\end{center}
\end{table}

Hard diffractive processes studied by CDF include $JJ$ (dijet), $W$, $b$-quark, and $J/\psi$ production. Results from Run~I have been published in Phys. Rev. Letters~(see review in Ref.~\cite{ref:lathuile}), and preliminary Run~II results have been presented at various conferences, including the present one (see Sec.~\ref{experiment}).

The most interesting aspects of the results in connection with the QCD structure of the diffractive exchange  are the breakdown of factorization and the restoration of factorization in events with multiple rapidity gaps.  The following two paragraphs are based on  excerpts from Sec.~\ref{experiment}, and are provided here for the convenience of the reader.
\paragraph{Breakdown of factorization.}
At $\sqrt s=$1800 GeV, the SD/ND ratios (gap fractions)   
for dijet, $W$, $b$-quark, and $J/\psi$ production, as well the ratio of
DD/ND dijet production, are all $\approx 1\%$.
This represents a suppression of $\sim {\cal{O}}(10)$
relative to predictions based on 
diffractive parton densities measured from Diffractive Deep Inelastic Scattering (DDIS) at HERA, indicating a breakdown of QCD factorization comparable to that
observed in soft diffraction processes relative to Regge theory expectations (see Sec.~\ref{intro}).  However, factorization approximately holds among the above four hard diffractive processes at fixed~\,$\sqrt s$, which suggests that the suppression is in the rapidity gap formation probability, as predicted by the generalized gap renormalization model (RENORM), which is the subject of this section. 

\paragraph{Restoration of factorization in multi-gap diffraction.}
Another interesting CDF result is that 
ratios of two-gap to one-gap cross sections 
for both soft and hard processes obey factorization. This provides not only  a clue to understanding diffraction in terms of a composite Pomeron, but also a potential experimental discovery tool for new physics using processes with 
multiple rapidity gaps.

\subsection{Renormalized diffractive cross sections: soft diffraction}

Diffraction has traditionally been treated in Regge theory
using factorization. This approach was successful at $\sqrt s$ 
energies below $\sim 50$ GeV~\cite{physrep}, but as the available energies 
increased to $\sqrt s=$1800 GeV in Run~I at the Tevatron, 
a suppression as large as $\sim {\cal{O}}(10)$
of the SD cross section was observed relative to the Regge theory based predictions~\cite{sd}.
This breakdown of factorization was traced to the energy dependence of 
$\sigma_{sd}^{tot}(s)\sim s^{2\epsilon}$, 
which is faster than that of $\sigma^{tot}(s)\sim s^\epsilon$, 
so that at high $\sqrt s$ unitarity would have to be 
violated if factorization held. 
The $s$-dependence appears explicitly in the SD differential 
cross section:
\begin{equation}
\hbox{ Regge theory: }
d\sigma_{sd}(s,M^2)/dM^2\sim \frac{s^{2\epsilon}}{(M^2)^{1+\epsilon}}.
\label{eq:reggeM2}
\end{equation}

As seen in Fig.~\ref{fig:rapgap}~{\em (right)}, contrary to the Regge theory based expectation of Eq.~(\ref{eq:reggeM2}), the
measured SD $M^2$-distribution does not show any 
$s$-dependence over a region of $s$ six orders of 
magnitude. Thus, it appears that factorization breaks down 
in such a way as to enforce $M^{\;2}$-scaling. This property is built into the RENORM model, in which the 
Regge theory Pomeron flux is renormalized to unity~\cite{R}. 
Below, we present a QCD basis for renormalization 
and its extension to central and multi-gap diffraction~\cite{corfu}.

The form of the rise of total cross sections at high energies, 
$\sim s^{\;\epsilon}$, which in Regge theory requires a Pomeron trajectory 
with intercept $\alpha(0)=1+\epsilon$, is expected  
in a parton model approach, where cross sections are 
proportional to the number of available wee partons~\cite{Levin}.
In terms of the rapidity region in which there is particle 
production~\footnote{We take $p_T=1$ GeV so that $\Delta y'=\Delta \eta'$.}, 
$\Delta \eta'$, the total $pp$ cross section is given by
\begin{equation}
\sigma_{pp}^{tot}=\sigma_0\cdot e^{\epsilon\Delta\eta'}.
\label{totDeta}
\end{equation}
Since from the optical theorem $\sigma_{tot}\sim {\rm Im\,f^{\;el}}(t=0)$, 
the full parton model amplitude takes the form 
\begin{equation}
{\rm Im\,f^{el}}(t,\Delta\eta)\sim e^{({\epsilon}+\alpha't)\Delta \eta},
\label{eq:fPM}
\end{equation}
\noindent where the term  $\alpha't$ is 
a parameterization of the $t$-dependence of the amplitude. 
Based on this amplitude, the diffractive cross sections of table~1
are expected to have the forms 
\begin{eqnarray}
\frac{d^2\sigma_{sd}}{dt\,d\Delta\eta}=N^{-1}_{gap}(s)& 
F_p(t)\left\{e^{[\epsilon+\alpha'(t)]\Delta\eta}\right\}^2
 &\kappa\left[\sigma_0e^{\epsilon\Delta\eta'}\right]
\nonumber\\
\frac{d^3\sigma_{dd}}{dt\,d\Delta \eta\,d\eta_c}=N^{-1}_{gap}(s)&
\left\{e^{[\epsilon+\alpha'(t)]\Delta\eta}\right\}^2
&\kappa\left[\sigma_0e^{\epsilon(\Sigma_i\Delta\eta_i')}\right]
\nonumber\\
\frac{d^4\sigma_{sdd}}{dt_1\,dt_2\,d\Delta \eta\,d\eta_c}=N^{-1}_{gap}(s)&
F_p(t)\Pi_i\left\{e^{[\epsilon+\alpha'(t_i)]\Delta\eta_i}\right\}^2
&\kappa^2\left[\sigma_0e^{\epsilon(\Sigma_i\Delta\eta_i')}\right]
\nonumber\\ 
\frac{d^4\sigma_{dpe}}{dt_1\,dt_2\,d\Delta \eta\,d\eta'_c}=N^{-1}_{gap}(s)&
\underbrace{\Pi_{i}\left\{F_p(t_i)e^{[\epsilon+\alpha'(t_i)]\Delta\eta_i}\right\}^2}_{\hbox{gap probability}} 
&\kappa^2\underbrace{\left[\sigma_0e^{\epsilon(\Delta\eta')}\right]}_{\textstyle{\sigma^{tot}(s')}},
\label{eq:diffPM}
\end{eqnarray}
where the (re)normalization factor 
$N_{gap}(s)$ is the integral of the gap probability over all 
phase space in ($t_i$, $\Delta\eta_i$, $\eta_c$, $\eta'_c$), and the variables $\eta_c$ and $\eta'_c$ represent the center of the floating (not adjacent to 
a nucleon) rapidity gap in DD or SDD and the floating diffractive cluster in DPE, respectively. In each case, the independent variables are the ones 
on the left hand side of the equation, but for pedagogical reason  we 
use on the right side the additional variables $\Delta\eta'_i$, which could be be expressed in terms of $\ln s$ and the variables on the left. 

The expressions in Eq.~(\ref{eq:diffPM}) are built from the following components:
\begin{itemize}
\item {\em the reduced energy cross section}, $\sigma^{tot}_{pp/p\bar p}(s')$, which is the  $pp/p\bar p$ cross section at the reduced collision energy, which is defined by the equation $\ln (s'/s_0)=\sum_i\Delta y_i'$; 
\item {\em the color factors $\kappa$}, one for each gap, 
required to select color neutral exchanges with vacuum quantum numbers to ensure diffractive rapidity gap formation;
\item {\em the gap probability}, which is given by the amplitude squared of the elastic scattering 
between a diffractively dissociated and a surviving proton, in which case 
it contains the proton form factor $F_p(t)$, or between two diffractively 
dissociated protons;
\item {\em the normalization factor $N^{-1}_{gap}$}, which is the inverse of the integral of the gap probability over all phase space.
\end{itemize}
A remarkable property of the expressions in Eq.~(\ref{eq:diffPM}) is that they factorize 
into two terms, one which depends on the sum of the rapidity regions in which there is 
particle production, and another which  depends on the sum of the widths of the rapidity gaps. 
This is rendered possible by the exponential dependence 
on $\Delta\eta$ of the elastic amplitude, which allows non-contiguous 
regions in rapidity to be added in the exponent. A consequence of this
type of factorization is that the (re)normalization factor is the same and is $\sim s^{2\epsilon}$ in all cases, 
ensuring $M^2$-scaling and universality of the suppression factor across single, central, and multi-gap diffraction.

\begin{figure}[htp]
{\psfig{figure=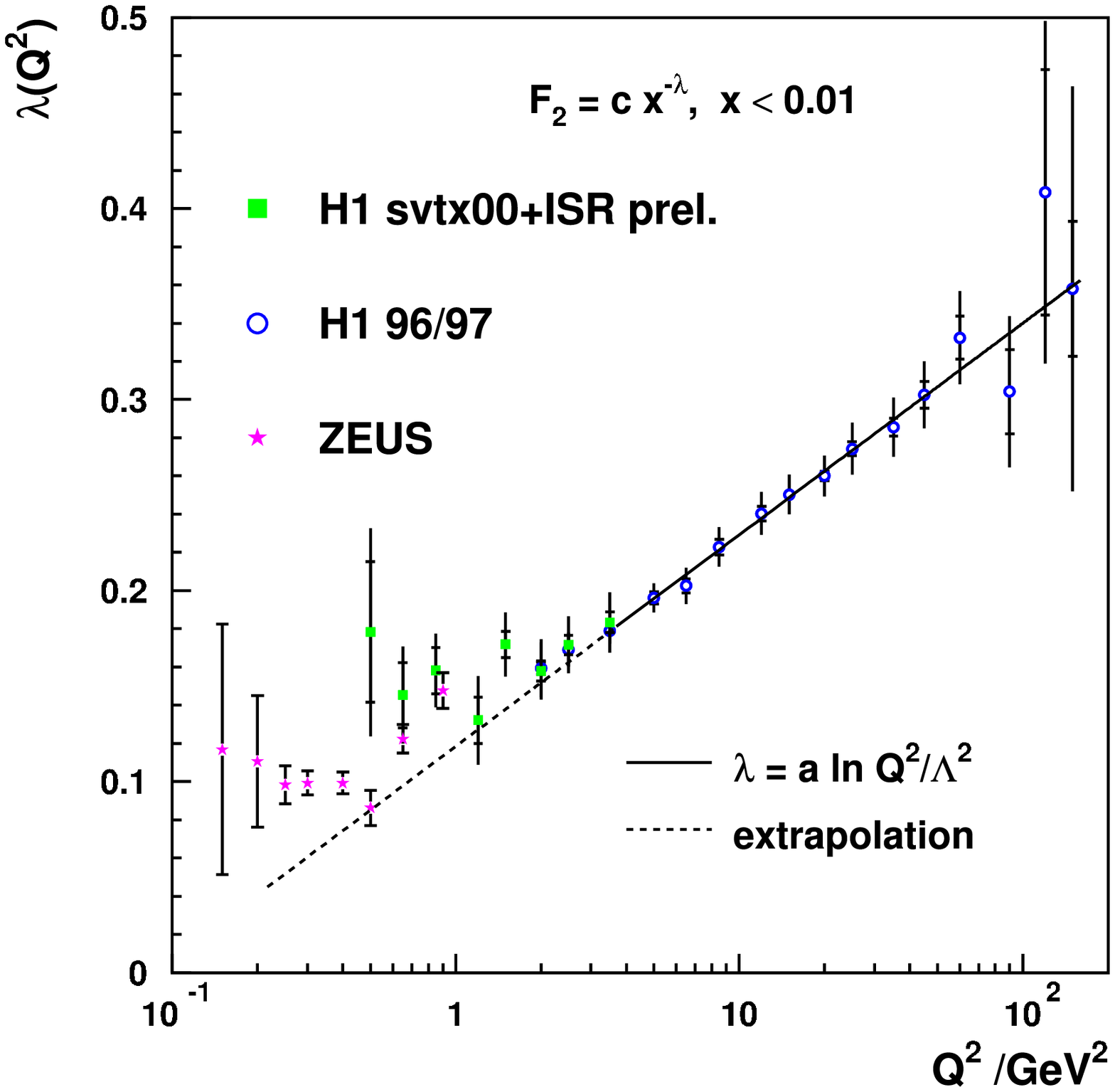,width=0.5\textwidth}}
\vspace*{-22.5em}
{\hspace*{0.53\textwidth}\psfig{figure=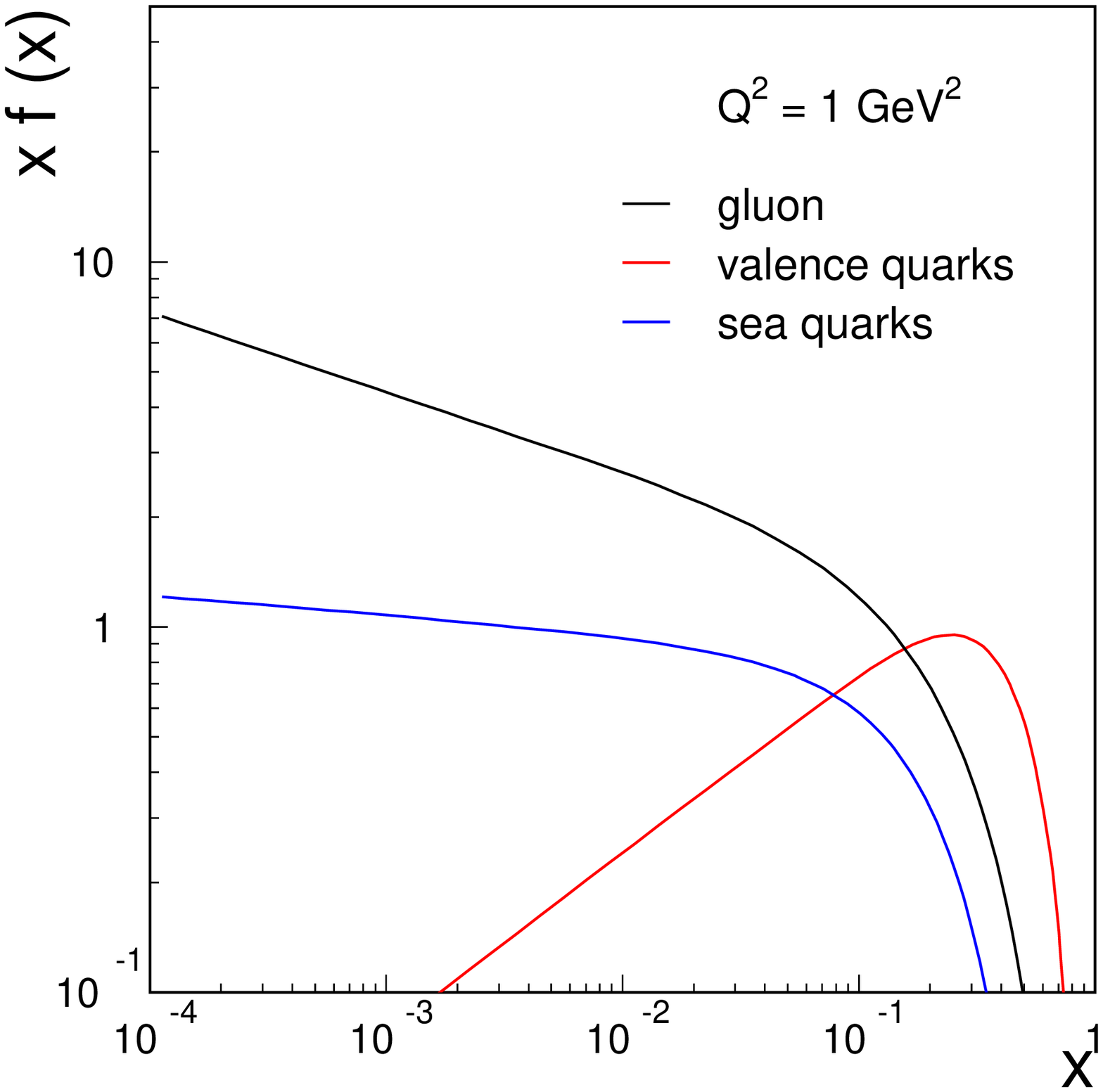,width=0.5\textwidth}}
\caption{\it  ({left}) The parameter $\lambda(Q^2)$ vs. $Q^2$ 
of a fit to the structure function $F_2(x,Q^2)\sim x^{-\lambda(Q^2)}$ in DIS at HERA\protect\cite{heraglue};
({right}) CTEQ5L nucleon parton distribution functions for $Q^2=1$ GeV$^2$.} 
\label{fig:softQCD}
\end{figure}

\paragraph {The parameters $\textstyle\epsilon$ and $\textstyle\kappa$.}
Experimentally, these parameters have been 
measured to be~\cite{CMG,GM}
\begin{eqnarray}
\epsilon\equiv &\alpha_{\pom}(0)-1&=0.104\pm 0.002\pm 0.01\hbox{ (syst)}, \mbox{ and}\nonumber\\
\kappa\equiv &\frac{\textstyle{g}_{\pom\pom\pom}}{\beta_{\pom p}}&=0.17\pm0.02\hbox{ (syst)},
\label{eq:ek_exp}
\end{eqnarray} 
where the systematic uncertainty assigned to $\epsilon$ is an 
estimate  
based on considering results from fits made to cross section data by various authors. 
Measurements of parton densities at HERA indicate that partonic 
structure in the nucleon is expressed down to the 
hadron mass scale of $Q^2\approx 1$ GeV$^2$. 
This is seen in Fig.~\ref{fig:softQCD}~({\em left}), where the
parameter $\lambda(Q^2)$ of $F_2(x, Q^2)\sim x^{-\lambda(Q^2)}$   
decreases linearly with $\ln Q^2$ down to $Q^2\approx 1$ GeV$^2$, 
flattening out and becoming consistent with the soft Pomeron intercept only below 
$Q^2=1$ GeV$^2$. We therefore assume  partonic structure in soft diffractive exchanges at the hadron mass scale, and proceed with a ``toy estimate'' of the parameters $\epsilon$ and
$\kappa$ from the nucleon PDF, using the  PDF 
at $Q^2=1$ GeV$^2$ shown in Fig.~\ref{fig:softQCD}~({\em right})  
obtained from the CTEQ5L parameterization. 

%
%
%
The region of interest to diffraction, $x\leq 0.1$, is dominated by sea
gluons and quarks.
In this region, a fit of the form $xf(x)\sim x^{-\lambda}$, Fig.~\ref{fig:softQCD}~({\em right}), yields 
$\lambda_g\approx 0.2$ and $\lambda_q\approx 0.04$ with relative weights 
$w_g\approx 0.75$ and $w_q\approx 0.25$~\footnote{For valence quarks, $\lambda_q\equiv \lambda_R\approx -0.5$; this is relevant for Reggeon exchange, which is not being considered here, as its contribution is relatively small.}.
Noting that 
the number of wee partons grows as $\int_{1/s}^1 f(x)dx\sim s^\lambda$, 
the Pomeron intercept may be obtained from the parameters $\lambda_g$ and 
$\lambda_q$ appropriately weighted by the 
gluon and quark color factors
\begin{equation}
c_g=\frac{1}{N_c^2-1},\;\;\;\;c_q=\frac{1}{N_c}.
\label{eq:color}
\end{equation}       
The weighting procedure places $\epsilon$ in the range $\lambda_q<\epsilon<\lambda_g$, or $0.04<\epsilon<0.2$, 
which covers the experimental value of $\epsilon=0.104$. A precise determination is not attempted, as it would require averaging over the $Q^2$ range of all the particles produced in the collision and proper accounting of the uncertainties in the nucleon PDF in this low $Q^2$ region.    

The  parameter $\kappa$ is obtained from the gluon and quark color factors and weights:
\begin{eqnarray}
\kappa\approx &c_gw_g+c_qw_q&=0.182.
\label{eq:k_th}
\end{eqnarray}
This RENORM prediction is in remarkably good agreement with 
$\kappa_{exp}=0.17\pm0.02$.  

\subsection{Renormalized diffractive cross sections: hard diffraction}
Hard diffraction processes are defined as those in which there is a hard 
partonic scattering in addition to the diffractive rapidity gap 
signature. Events may have forward, central, or multiple rapidity 
gaps in topologies similar to those listed in table~1 for soft $\bar pp$ collisions at the Tevatron, with the hard scattering products appearing within the region(s) of rapidity where there is particle production.  
 
\paragraph{Hard diffraction data.}  
CDF has measured SD/ND ratios for $W$, dijet, $b$-quark and $J/\psi$ 
production, and also diffractive structure functions extracted from SD and DPE dijet production (see Sec.~\ref{experiment}. 
HERA has reported DPFs extracted from inclusivs DDIS, as well as from exclusive channels~\cite{HERA_this_conference}.

The following interesting characteristics have been observed:
\begin{itemize}
\item all SD/ND ratios measured by CDF at $\sqrt s=$1800 GeV are 
approximately equal, pointing to a  flavor independent
rapidity gap formation probability;
\item the dijet SD/ND ratio measured by CDF varies as $\sim x_{Bj}^{-0.45}$, contrary to results from deep inelastic scattering at HERA, where a constant ratio of DDIS/DIS is observed~\cite{heraglue,HERA_this_conference};
\item the SD structure function extracted from dijet production at CDF 
is suppressed by $\sim{\cal{O}}(10)$ relative to expectations 
from diffractive PDFs measured in 
DDIS at HERA; 
\item the Pomeron intercept measured in DDIS at HERA
increases with $Q^2$ and is on average larger 
than the soft Pomeron intercept, but approximately a factor of $\sim 2$ smaller than the intercept obtained from inclusive DIS.
\end{itemize} 

\paragraph{Diffractive parton densities.} Diffractive parton densities with good statistics have been obtained from DDIS at HERA and diffractive dijet production at the Tevatron: 
\begin{eqnarray}
\hbox{ HERA:}&&\gamma^*+p\rightarrow p+jet+X\\
\hbox{ Tevatron:}&&\stackrel{-}{p}+p\rightarrow \stackrel{-}{p}+\hbox{dijet}+X.\nonumber
\label{SD_hera_Tev}
\end{eqnarray}
The production process may involve several color ``emissions''
from the surviving proton, collectively comprising a color singlet with vacuum quantum numbers. Two of the emissions are of special importance: 
one at $x=x_{Bj}$ from the proton's PDF at scale $Q^2$, which is responsible 
for the hard scattering, and another at $x=\xi$ 
(fractional momentum loss of the diffracted nucleon) from the PDF at scale $Q^2\approx 1$ GeV$^2$, which neutralizes the exchanged color 
and forms the rapidity gap. 

\paragraph{At HERA,} at small $\xi$ where the proton PDF exhibits power law behavior at both soft and hard scales, the diffractive structure function takes the form~\footnote{For simplicity, 
we do not consider the $t$ dependence in this discussion. This has little affect on our conclusions, as diffractive cross sections are concentrated at small $t$.}
\begin{equation}
F_2^{D(3)}(\xi,x,Q^2)=\frac{A_{\rm norm}}{\xi^{1+\textstyle{\epsilon}_q}}\cdot c_q\cdot F_2(x,Q^2),
\label{eq:F2D3}
\end{equation}
where $A_{\rm norm}$ is a normalization factor, $\epsilon_q$ is the exponent of the soft quark structure function, which from Fig.~\ref{fig:softQCD}~({\em right\,}) is given by $\epsilon_=\lambda_q(Q^2=1)=0.04$, and $c_q=1/3$ is the quark color factor.

\begin{figure}[htp]
\vspace*{2cm}
\underline{\large Definition of $\lambda$}:\\
{\Large $x\cdot f(x)\propto x^{-\lambda}$}
\vfill
\vspace*{-4cm}
{\hspace*{15em}\psfig{figure=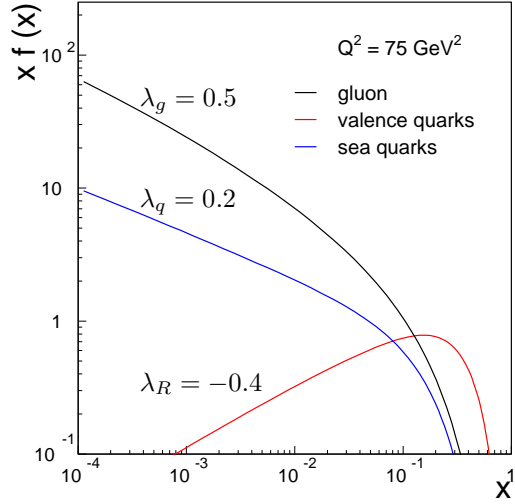,width=0.6\textwidth}}

\vspace*{-5.7cm}
\hspace*{20em}$\lambda_g=0.5$\\

\vspace*{0.5cm}
\hspace*{20em}$\lambda_q=0.2$\\

\vspace*{1.6cm}
\hspace*{20em}$\lambda_R=-0.4$\\

\vspace*{1cm}
\caption{\it CTEQ5L nucleon parton distribution functions for $Q^2=75$ GeV$^2$. The parameters $\lambda_{g,q,R}$ are the slopes of the gluon, 
sea quark, and valence quark distribution (`$R$' stands for $Reggeon$) 
in the region of $x<0.1$, where the power law 
behavior holds.} 
\label{fig:hardQCD}
\end{figure}

At high $Q^2$, where factorization is expected to 
hold~\cite{R,JCollins}, $A_{\rm norm}$ is the normalization factor of the soft PDF, which is a constant, and $F_2$ can be expressed as a power law, resulting in
\begin{equation}
F_2^{D(3)}(\xi,x,Q^2)=\frac{A_{\rm norm}}{\xi^{1+\textstyle{\epsilon}_q}}\cdot \frac{1}{3}\cdot
\frac{C(Q^2)}{(\beta\xi)^{\lambda(Q^2)}}
=\frac{A_{\rm norm}}{\xi^{1+\textstyle{\epsilon_q}+\lambda(Q^2)}}\cdot \frac{1}{3}\cdot
\frac{C(Q^2)}{\beta^{\lambda(Q^2)}},
\label{eq:F2D3hiQ}
\end{equation}
where $\lambda(Q^2)$ is the power of the fit of inclusive data to the form $F_2(x,Q^2)\sim x^{-\lambda(Q^2)}$ shown in Fig.~\ref{fig:softQCD}~{\em(left\,)}.

The expression in Eq.~(\ref{eq:F2D3hiQ}) leads to two 
important RENORM  predictions:
\begin{itemize}
\item the Pomeron intercept in DDIS is the average of the soft quark and inclusive interceps,  
\begin{equation}
\alpha^{DIS}_{\pom}=1+\lambda(Q^2),\;\;\;\;\;
\alpha^{DDIS}_{\pom}=1+\frac{1}{2}\left[\epsilon_q+\lambda(Q^2)\right];
\label{eq:heraintercept}
\end{equation}
\item the ratio of DDIS to DIS structure functions at fixed $\xi$ 
is independent of $x$ and $Q^2$,
\begin{equation}
R\left[\frac{F^{D3}_2(\xi,x,Q^2)}{F_2(x,Q^2)}\right]_{\xi}=\frac{1}{3}\cdot
\frac{A_{\rm norm}}{\xi^{1+\epsilon}}.
\label{eq:Rhera}
\end{equation}     
\end{itemize}
Data from ZEUS and H1 are consistent with these predictions~\cite{heraglue}.

At low $Q^{\;2}$, the gap probability saturates and renormalization must be applied~\cite{R}. In the RENORM model, the resulting suppression factor depends on the size of the rapidity interval available for particle production, which is $\Delta y'=\ln s-\ln Q^2=\ln ({s}/{Q^2})$. For dijet photoproduction, where $\Delta y'\approx 6$, a suppression factor of $\approx 3$ is expected by RENORM~\cite{R}, similar to that observed in soft SD at the Tevatron. 
Data from dijet photoproduction at HERA are consistent with this prediction~\cite{HERA_this_conference}. Moreover, since the suppression in RENORM is due to saturation of the gap probability, the same suppression is expected for both direct and resolved rates at low $Q^{\,2}$ DDIS. This prediction is also consistent with observation~\cite{HERA_this_conference}.  

\paragraph{At the Tevatron,} where the gap probability saturates and must be renormalized to unity, the RENORM diffractive structure function takes the form
\begin{eqnarray}
F_{JJ}^{D3}(\xi,x,Q^2)= N_{\rm gap}^{-1}(s,\beta)\cdot \frac{1}{\xi^{1+2\epsilon}} \cdot F_{JJ}(\frac{x}{\xi},Q^2)\\
N_{\rm gap}(s,\beta)=\int^{\xi=0.1}_{\xi_{mim}}\frac{d\xi}{\xi^{1+2\epsilon}}\approx \frac{(\beta s)^{2\epsilon}}{2\epsilon},
\label{ref:TevDSF} 
\end{eqnarray}
where, as in DDIS, $\epsilon$ is the power from the fit of the soft structure function to the form $x\,f(x)\sim {x}^{-\epsilon}$, and the limits $\xi_{min}=x_{min}/\beta$ and $x_{min}=1/s$ are used~\cite{R}. Through renormalization, $F_{JJ}^{D3}$ acquires a factor $\sim (1/\beta)^{2\epsilon}$, and the diffractive to inclusive structure function ratio, $R_{JJ}(SD/ND)$, a factor $\sim (1/x)^{2\epsilon}$.  Since from color factor considerations $F_{JJ}$ is gluon dominated,
\begin{equation}
F_{JJ}(x)=x\left[g(x)+\frac{4}{9}q(x)\right],
\label{eq:SF}
\end{equation}
where $g(x)$ and $q(x)$ are the gluon and quark densities in the proton, the relevant  parameter $\epsilon$ is the parameter $\lambda_g$ of Fig.~\ref{fig:softQCD}~{\em(right)}, resulting in $R_{JJ}(SD/ND)\sim 1/x^{0.4}$. 
This RENORM prediction is confirmed by the CDF data,
where the $x$-dependence of the diffractive to inclusive ratio 
was measured to be $\sim 1/x^{\;0.45\pm0.02}$~\cite{dijet1800}. 

\subsection{Summary of phenomenology section}
A phenomenological model has been presented (RENORM), in which diffractive cross sections are obtained from parton-level cross sections and the underlying inclusive parton distribution function of the interacting hadrons using QCD color factors and appropriate (re)normalization. 
Scaling and factorization properties observed in fixed target, Tevatron, and HERA data form both the basis and a testing ground for RENORM. 

In soft diffraction, normalized single-gap and double-gap differential cross sections (SD, DD, DPE, SDD) are obtained in RENORM in terms of two free parameters, $\epsilon$ and $\alpha'$, which are identified as those of the Regge theory soft Pomeron trajectory, $\alpha(t)=1+\epsilon+\alpha't$. Furthermore, the $\epsilon$ is obtained from the color-factor weighted exponents of power law fits to soft nucleon PDFs, leaving $\alpha'$ as the only free parameter in the model.

In hard diffraction, RENORM is applied to HERA and Tevatron data.
At HERA, interesting features of the data include: $\epsilon(Q^2)_{DDIS}<\epsilon(Q^2)_{DIS}$,  $F^{D3}_2(\xi,x,Q^2)/F_2(x,Q^2)|_\xi\sim  \mbox{ constant}$,   dijet photoproduction suppressed by factor of $\sim 3$, and  direct~/~resolved low $Q^2$ DIS both suppressed by approximately the same factor relative to high $Q^2$ DIS. 
At the Tevatron, features of the data include: cross sections at fixed collision energy are flavor independent, the ratio of SD/ND rates decreases with increasing $x_{Bj}$,  and the DSF of the proton in DPE events with a leading $\bar p$ is suppressed relative to that in SD. Comparing HERA with Tevatron results, factorization breaks down at the Tevetron relative to HERA by a factor of $\sim {\cal{O}}(10)$, but is restored in comparing the double-gap DSF obtained from DPE at CDF with that obtained from DDIS of HERA. 
All these  features are successfully interpreted by RENORM.

\section{Conclusion} In light of the success of the parton-model based renormalization (RENORM) approach in describing the data, diffraction may be viewed as  an interaction between low-$x$ partons subject to color constraints.
%
%

\end{document}